\documentclass[a4paper,10pt]{article}

\usepackage{amsfonts}
\usepackage{amsmath}
\usepackage[all]{xy}
\usepackage{verbatim}
\usepackage{rotating}

\usepackage[ps2pdf]{hyperref}
\hypersetup{pdftitle = {Fixed Point Resolution in Extensions of Permutation Orbifolds}}
\hypersetup{pdfauthor = {Michele Maio, Bert Schellekens}}
\hypersetup{pdfsubject = {CFT-String Theory}}
\hypersetup{pdfkeywords = {Orbifolds, Extensions, Fixed Point Resolution}}

\numberwithin{equation}{section}
\begin{document}

\title{Fixed Point Resolution in Extensions of Permutation Orbifolds}
\author{M. Maio$^{1}$ and A.N. Schellekens$^{1,2,3}$\\~\\~\\
\\
$^1$Nikhef Theory Group,
Amsterdam, The Netherlands
\\
~\\
$^2$IMAPP, Radboud Universiteit Nijmegen, The Netherlands
\\
~\\
$^3$Instituto de F\'\i sica Fundamental, CSIC, Madrid, Spain}

\maketitle

\begin{abstract}
We determine the simple currents and fixed points of the orbifold theory $CFT\otimes CFT/\mathbb{Z}_2$, given the simple currents and fixed point of the original $CFT$. We see in detail how this works for the $SU(2)_k$ WZW model, focusing on the field content (i.e. $h$-spectrum of the primary fields) of the theory. We also look at the fixed point resolution of the simple current extended orbifold theory and determine the $S^J$ matrices associated to each simple current for $SU(2)_2$ and for the $B(n)_1$ and $D(n)_1$ series.
\end{abstract}

\clearpage
\tableofcontents

\section{Introduction}
Conformal field theories \cite{Belavin:1984vu} are well established tools in String Theory. Orbifold conformal field theories are standard way of deriving new conformal fields theories out of existing ones. The prototype example is a  of the form $G/H$ where, after moding out the symmetry group $H$, one is left with  $H$-invariant states plus twisted fields, necessary to ensure modular invariance. Orbifold CFT's appear all over the place in string theory, for example in Gepner models \cite{Gepner:1987vz,Gepner:1987qi} where one builds the ``internal'' sector of string compactifications out of tensor products of $N=2$ minimal models with total conformal anomaly $c=\sum_i c_i=9$.

In this paper we will consider orbifolds by cyclic permutations of tensor product conformal field theories. We  start with a given CFT, take the tensor product of $\lambda$ copies of it and mod out by the cyclic symmetry $Z_\lambda$, which generates the full permutation group $S_\lambda$. The field content of such cyclic orbifold theories was worked out already long ago by Klemm and Schmidt \cite{Klemm:1990df} who were able to read off the twisted fields using modular invariance. Later, Borisov, Halpen and Schweigert \cite{Borisov:1997nc} introduced an orbifold induction procedure, providing a systematic construction of cyclic orbifolds, including their twisted sector, and determining orbifold characters and, in the $\lambda=2$ case, their modular transformation properties. Generalizations to arbitrary permutation groups were done by Bantay \cite{Bantay:1997ek,Bantay:1999us}.

Extensions with integer spin simple currents \cite{Schellekens:1989am,Intriligator:1989zw} are essential tools in conformal field theories (see \cite{Schellekens:1990xy} for a review). In string theory, they appear when it is needed to make projections (e.g. GSO projection) or implement constraints (such as world-sheet supersymmetry constraints, or
the so-called $\beta$-constraints in Gepner models, which impose world-sheet and space-time supersymmetry). Simple current extensions are also used to implement field identification in coset models.
In extensions, some fields of the original theory are projected out while the remaining ones organize themselves into orbits of the current. Fixed points are very particular orbits, those with length one; more generally if a simple current discrete group
has an order that is not prime, there can be orbits whose length is a divisor of the order, due to fixed points of powers of
the generator of the group.
In presence of fixed points, extensions are much more complicated to handle, since each of them give rise to a number of ``splitted'' fields on which there is a priori no control. In particular, determining the $S$ matrix of the extended theory is straightforward if there are no fixed points, but it is highly non-trivial otherwise. In presence of fixed points, the knowledge of the full $S$ matrix is parametrized by a set of  ``$S^J$'' matrices \cite{Fuchs:1996dd}, one for each simple current $J$: knowing all the $S^J$ matrices amounts to knowing the $S$ matrix of the extended theory. Fixed points can also appear for
half-integer spin currents, and the corresponding matrices $S^J$ are important when these currents are combined to form
integer spin currents. Furthermore, simple current fixed points and their resolution matrices are essential ingredients for
determining the boundary coefficients in a large class of rational CFT's \cite{Bianchi:1991rd,Fuchs:2000cm}.

The determination of fixed point matrices $S^J$ was first considered in \cite{Schellekens:1989uf}. There an empirical
approach was used, based on the information that these matrices must satisfy modular group properties. Hence an {\it ansatz} could
be guessed in some simple cases from the known fixed point spectrum. These {\it ans\"atze} were proved and extended in 
\cite{Fuchs:1995zr}, where they were related to foldings of Dynkin diagrams. Starting from these results,
the $S^J$ matrices are now known in many cases, such as for WZW models \cite{Schellekens:1990xy,Schellekens:1999yg} and coset models \cite{Schellekens:1989uf}. Here we would like to determine the set of $S^J$ matrices for cyclic permutation orbifolds. In this paper we restrict ourselves to $Z_2$ permutation orbifolds of an original CFT. We will manage to determine the $S^J$ matrices in a few, but interesting, cases, namely for the $SU(2)_2$ WZW and for the $B(n)_1$ and $D(n)_1$ series. The method we use is
based on the fact that the extensions corresponding to these cases are CFT's whose matrix $S$ can also be obtained by
other means. However, even though the matrices $S^J$ are not needed to construct the matrix $S$ of the
extension, the result still provides important new information, because these matrices can also be used in tensor products
and coset CFT's for combinations of currents. A case that is
particularly interesting is the application to $N=2$ minimal models, where $SO(2)_1$ factor appears in the definition 
in terms of the coset CFT. We hope to return to this case in the future. We expect that the solution we present here for
an infinite series of special cases will provide insight in the general case, and leads to an {\it ansatz} that can be
checked. This is also left for future work.

The outline of the paper is as follows.\\
In section \ref{Definition of the problem} we define the problem that we would like to address, namely the resolution of the fixed points in extension of permutation orbifolds. Fixed points are problematic since basic CFT quantities, such as the $S$ matrix of the extended theory, cannot be easily derived. Moreover, they carry an intrinsic ambiguity, that in some case does not matter, e.g. when we look at the $S$ matrix of the extended theory, but in other situations is important and might change the result, e.g when we look at the set of $S^J$ matrices.\\
Before going into the details of the problem, in section \ref{Simple currents of the orbifold CFT} we study a bit more systematically the structure of simple currents and corresponding fixed points in orbifold CFT's. In particular, we will see which simple currents and fixed points can arise in the orbifold theory starting from simple currents and fixed points in the mother theory. This is an application of \cite{Borisov:1997nc}.\\
Section \ref{Example SU(2)k} provides an example where the mother theory is $SU(2)_k$.\\
Next we move to the main problem, i.e. the fixed point resolution in extension of permutation orbifolds. We present the results in section \ref{Fixed point resolution in SU(2)k orbifolds} and section \ref{Fixed point resolution in SO(N)1 orbifolds} for $SU(2)_1$ and $SO(N)_1$. We say something about arbitrary level $k$ as well.\\
In the appendix some extracts from calculations are given.\\
We do not solve the problem in full generality, i.e. for every value of $k$ and for any arbitrary mother CFT: this is left for future work.
In the first part of the paper we will mostly work with simple currents $J$ of order two, i.e. $J^2=2$, whose orbits can have length one or two. In the second part, higher-order currents will become important.

\section{Definition of the problem}
\label{Definition of the problem}
Given a certain CFT, we would like to look at the orbifold theory with $\lambda=2$: 
\begin{equation}
 (CFT)_{\rm perm} \equiv CFT\times CFT/\mathbb{Z}_2\,.
\end{equation}
Moding out by $\mathbb{Z}_2$ means that the spectrum must contain fields that are symmetric under the interchange of the two factors. This theory admits an untwisted and a twisted sector. The untwisted fields are those combinations of the original tensor product fields that are invariant under this flipping symmetry. Their weights are simply given by the sum of the two weights of each single factor. Twisted fields are required by modular invariance. In general, for any field $\phi_i$ in the original CFT, there are exactly $\lambda$ twisted fields in the orbifold theory, labelled by $\psi=0,1,\dots,\lambda-1$. Their weights were derived in \cite{Klemm:1990df} and are given by
\begin{equation}
\label{KS weight for twisted fields}
 h_{\widehat{(i,\psi)}}=\frac{h_i}{2}+\frac{c}{24}\frac{(\lambda^2-1)}{\lambda}+\frac{\psi}{2}\,,
\end{equation}
where $h_i\equiv h_{\phi_i}$ and $c$ is the central charge of the original CFT.

If there is any integer or half-integer spin simple current in the original CFT, it gives rise to an integer spin simple current in the orbifold CFT, which can be used to extend the orbifold CFT itself. In the extension, some fields are projected out while the remaining organize themselves into orbits of the current. Typically untwisted and twisted fields do not mix among themselves. As far as the new spectrum is concerned, we do know that these orbits become the new fields of the extended orbifold CFT, but we do not normally know the new $S$ matrix. From now on we will write $\tilde{S}$ with a tilde to denote the $S$ matrix of the extended theory.

If there are no fixed points, i.e. orbits of length one, the $S$ matrix of the extended theory, $\tilde{S}$, is simply given by the $S$ matrix of the unextended theory (in case of permutation orbifolds it is the BHS $S$ matrix given in \cite{Borisov:1997nc}) multiplied by the order of the extending simple current. Unfortunately, often this is not the case: normally there will be fixed points and the extended $S$ matrix cannot be easily determined.

Using the formalism developed in \cite{Fuchs:1996dd}, we can trade our ignorance about $\tilde{S}$ with a set of matrices $S^J$, one for every simple current $J$, according to the formula
\begin{equation}
\label{main formula for f.p. resolution}
\tilde{S}_{(a,i)(b,j)}=\frac{|G|}{\sqrt{|U_a||S_a||U_b||S_b|}}\sum_{J\in G}\Psi_i(J) S^J_{ab} \Psi_j(J)^{\star}\,,
\end{equation}
These $S^J_{ab}$'s are non-zero only if both $a$ and $b$ are fixed points. This equation can be viewed as a Fourier transform and the $S^J$'s as Fourier coefficients of $\tilde{S}$. The prefactor is a group theoretical factor acting as a normalization and the $\Psi_i(J)$'s are the group characters acting as phases. In our calculations, where all the simple currents have order two, the normalization prefactor is $1/2$ and the group characters are just signs.

In this way, the problem of finding $\tilde{S}$ is equivalent to the problem of finding the set of matrices $S^J$. In this paper we want to address exactly this problem, but in the case of permutation orbifolds. Suppose we know (and we do!) the $S$ matrix of the orbifold theory, then extend it by any of its simple currents; what is the matrix $\tilde{S}$ of the new extended theory? Equivalently, given the fact that there will be fixed points in the extension, what are the matrices $S^J$ for all the integer spin simple currents $J$? Hence, we are dealing with the fixed point resolution in extensions of permutation orbifolds.

\section{Simple currents of the orbifold CFT}
\label{Simple currents of the orbifold CFT}
Consider a CFT which admits a set of integer-spin simple currents $J$. This means that the $S$ matrix satisfies the sufficient and necessary condition \cite{Dijkgraaf:1988tf} $S_{J0}=S_{00}$, where $0$ denotes the identity field of the CFT. Every CFT has at least one simple current, namely the identity. Here we would like to determine the simple currents of the orbifold theory $CFT\otimes CFT /\mathbb{Z}_2$. The only thing we need is the orbifold $S$ matrix given by BHS \cite{Borisov:1997nc}. Remember that the identity field of the orbifold theory is the symmetric representation of the identity ``$0$'' of the original CFT, here denoted by $(0,0)$.

It is probably useful to recall the BHS $S$ matrix. The convention for the orbifold fields is as follows. Orbifold twisted fields carry a hat: $\widehat{(i,\psi)}$; off-diagonal fields are denoted by $(ij)$ with $i\neq j$, diagonal fields by $(i,\psi)$. Here $i,j$ are fields of the mother theory and $\psi=0,\dots,\lambda-1$. The orbifold $S$ matrix for $\lambda=2$ is then \cite{Borisov:1997nc}:
\begin{eqnarray}
\label{BHS off-diag}
S_{(ij)(pq)}&=&S_{ip}\,S_{jq}+S_{iq}\,S_{jp} \nonumber\\
S_{(ij)(p,\psi)}&=&S_{ip}\,S_{jp} \nonumber\\
S_{(ij)\widehat{(p,\psi)}}&=&0
\end{eqnarray}
\begin{eqnarray}
\label{BHS diag}
S_{(i,\psi)(j,\chi)}&=&\frac{1}{2}\,S_{ij}\,S_{ij} \nonumber\\
S_{(i,\psi)\widehat{(p,\chi)}}&=&\frac{1}{2}\,e^{2\pi i\psi/2} \,S_{ip}
\end{eqnarray}
\begin{eqnarray}
\label{BHS twisted}
S_{\widehat{(p,\psi)}\widehat{(q,\chi)}}&=&\frac{1}{2}\,e^{2\pi i(\psi+\chi)/2} \,P_{ip}
\end{eqnarray}
where the $P$ matrix is defined by $P=\sqrt{T}ST^2S\sqrt{T}$. Sometimes we will write $S^{BHS}$ to refer to the orbifold $S$ matrix.

\subsection{Simple current structure}
Let us start with the off-diagonal fields of the orbifold and ask if any of them can be a simple current. If $i$ and $j$ are two arbitrary fields of the original CFT, denoting by $(i,j)$, with $i \neq j$, the corresponding off-diagonal field in the orbifold, in order for $(i,j)$ to be a simple current we have to demand for the $S$ matrix of the orbifold theory
\begin{equation}
 S_{(i,j)(0,0)}=S_{(0,0)(0,0)}
\end{equation}
which, upon using BHS formula, amounts to satisfying the constraint
\begin{equation}
 S_{i0} S_{j0}=\frac{1}{2} S_{00} S_{00}
\end{equation}
for the $S$ matrix of the original CFT. This relation is never satisfied because of the constraint $S_{i0}\geq S_{00}$, which holds for unitary CFT's. Consequently there are no simple currents coming from off-diagonal fields.

Let us do the same analysis for twisted fields. Twisted fields are denoted by $\widehat{(k,\psi)}$, where $k$ is a field in the original CFT and $\psi=0,\,1$. Now the constraint 
 \begin{equation}
 S_{\widehat{(k,\psi)}(0,0)}=S_{(0,0)(0,0)}
\end{equation}
would read
\begin{equation}
 \frac{1}{2}S_{k0}=\frac{1}{2} S_{00} S_{00}\,.
\end{equation}
This is also never satisfied, because of the same unitarity constraints as before. Once again there are no simple currents coming from twisted fields. 

Finally let us study the more interesting situation of diagonal fields as simple currents. A diagonal field is denoted by $(i,\psi)$, where $i$ is a field in the original CFT and $\psi=0,\,1$ corresponding to symmetric and anti-symmetric representation. Here the constraint
\begin{equation}
 S_{(i,\psi)(0,0)}=S_{(0,0)(0,0)}
\end{equation}
gives
\begin{equation}
 \frac{1}{2}S_{i0} S_{i0}=\frac{1}{2} S_{00} S_{00}\,,
\end{equation}
which is satisfied if and only if $i$ is a simple current. 

Hence we conclude that, despite the fact that the existence of simple currents in the orbifold theory is in general related to the $S$ matrix of the original CFT, there always exist definite simple currents in the orbifold theory: they are the symmetric and anti-symmetric representations of those diagonal fields corresponding to the simple currents of the original theory. In particular, since in the original CFT there is at least one simple current, namely the identity, in the orbifold theory there will be at least two, namely $(0,0)$ (trivial, because it plays the role of the identity) and $(0,1)$.

We will soon see that this pattern is respected for $SU(2)_k$ WZW models. They admit one integer-spin simple current (the identity) for $k$ odd and two (one of which is again the identity) integer-spin simple currents for $k$ even. Consequently, we will always find $(0,0)$ and $(0,1)$ as orbifold simple currents when $k$ is odd; when $k$ is even, there will be two additional ones denoted by $(k,0)$ and $(k,1)$.

\subsection{Fixed point structure}
Given our simple currents of the $CFT\otimes CFT /\mathbb{Z}_2$ theory, hereafter denoted by $(J,\psi)$ with $J$ a simple current of the original CFT, we now move on to study the structure of their fixed points. We start from the following general relation \cite{Schellekens:1989dq,Intriligator:1989zw} which holds for any simple current $J$:
\begin{equation}
\label{simple-current-fundamental-relation}
 \frac{S_{Ji}}{S_{0i}}=e^{{2 \pi i (h_J +h_i -h_{J\cdot i})}}\,.
\end{equation}
In the exponent, we recognize the monodromy charge $Q_J(i)$ of $i$ with respect to $J$:
\begin{equation}
Q_J(i)=h_J+h_i-h_{J\cdot i} \,\,\,\,{\rm mod}\,\,\mathbb{Z}\,.
\end{equation}
When $i=0$, this gives back the relation used above for integer spin simple currents: $S_{J0}=S_{00}$. Moreover fixed points $f$ of integer spin simple currents satisfy
\begin{equation}
\label{fixed-poinfs-of-integer-spin-simple-current}
 S_{Jf}=S_{0f}\,.
\end{equation}
Observe that this does not guarantee that if $f$ satisfies (\ref{fixed-poinfs-of-integer-spin-simple-current}) then $f$ is a fixed point\footnote{In fact, $h_J$ drops out from eq. (\ref{simple-current-fundamental-relation}) and we are left with $h_{J\cdot i}=h_i$ mod $1$.}. What might happen then is that we find additional solutions to this condition that might not be fixed points: so we might want to remove some of them. We will see later by looking at the fusion coefficients when this necessary condition is also sufficient. Nevertheless we will use here this relation to identify possible fixed points of integer spin simple currents:
\begin{equation}
 S_{(J,\psi)(\bullet,\bullet)}=S_{(0,0)(\bullet,\bullet)}\,.
\end{equation}
As we will see soon, this condition is in some cases necessary and sufficient to find fixed points. Fixed points can be untwisted (diagonal and/or off-diagonal) and/or twisted fields and we will find that only for diagonal fields a more careful analysis of the fusion coefficients is needed.

\subsubsection{Twisted sector}
Let us start by looking for fixed points coming from the twisted sector, since these are the easiest ones. The condition to be imposed is
\begin{equation}
 S_{(J,\psi)\widehat{(p,\chi)}}=S_{(0,0)\widehat{(p,\chi)}}\,.
\end{equation}
 After using BHS this reduces to 
\begin{equation}
\label{orbifold-twisted-fixed-poinfs-of-integer-spin-simple-current}
 \frac{1}{2}e^{2\pi i\psi/2}S_{Jp}=\frac{1}{2}S_{0p} 
\end{equation}
in terms of the $S$ matrix of the original CFT. Let us first notice that when $J$ is the identity, there is no news, since this constraint is either trivially satisfied (for $\psi=0$ all the twisted fields are fixed points of the identity) or impossible (for $\psi=1$ there are now fixed points coming from the twisted sector). When instead $J$ is not the identity, we find that $\widehat{(p,\chi)}$ is a fixed point of $(J,\psi)$ in the following cases (according to (\ref{simple-current-fundamental-relation})):
\begin{itemize}
 \item if $\psi=0$, when $p$ has integer monodromy charge with respect to $J$, i.e. $Q_J(p)=0$;
 \item if $\psi=1$, when $p$ has half-integer monodromy charge with respect to $J$, i.e. $Q_J(p)=\frac{1}{2}$.
\end{itemize}

\subsubsection{Off-diagonal fields}
Fixed points coming from off-diagonal fields must satisfy:
\begin{equation}
 S_{(J,\psi)(p,q)}=S_{(0,0)(p,q)}\,,
\end{equation}
with $p\neq q$. BHS then gives
\begin{equation}
 S_{Jp}S_{Jq}=S_{0p}S_{0q}\,.
\end{equation}
According to (\ref{fixed-poinfs-of-integer-spin-simple-current}), this is definitely satisfied if $p$ and $q$ are both fixed points of $J$. In particular, when $J$ is the identity, this relation is always true. Consequently, all possible off-diagonal fields $(p,q)$ are fixed points of the simple currents $(0,\psi)$. 

Focusing now on $J\neq 0$, other two possibilities, corresponding to different sign choices, are given by
\begin{equation}
\label{orbifold-offdiagonal-fixed-poinfs-of-integer-spin-simple-current}
\left\{
 \begin{array}{ccc}
   S_{Jp} &=& \pm S_{0q}\\ 
   S_{Jq} &=& \pm S_{0p}\,.
 \end{array}
\right.
\end{equation}
This is solved by $p$ and $q$ belonging to the same $J$-orbit, i.e. $p=Jq$, with either integer or half-integer monodromy charge with respect to $J$. In fact, 
\begin{equation}
 S_{Jp}=S_{J\cdot 0,J\cdot q}=e^{2\pi i Q_J(q)} e^{2\pi i Q_J(0)} e^{2\pi i Q_J(J)} S_{0q}= e^{2\pi i Q_J(q)} S_{0q}\,.
\end{equation}
However there might be more solutions.

\subsubsection{Diagonal fields}
For diagonal fixed points the condition is:
\begin{equation}
 S_{(J,\psi)(j,\chi)}=S_{(0,0)(j,\chi)}\,,
\end{equation}
which reduces to
\begin{equation}
 \frac{1}{2} (S_{Jj})^2=\frac{1}{2} (S_{0j})^2\,.
\end{equation}
after using BHS. The case $J=0$ is again trivial. Consequently, there are fixed points of $(0,\psi)$ for every $j$ of the original CFT. For $J\neq 0$, using (\ref{simple-current-fundamental-relation}), we see that $S_{Jj}=e^{2\pi i Q_J(j)}S_{0j}$, hence this constraint is solved by all fields $j$ with integer or half-integer monodromy charge. In particular, they include fixed points of $J$.

\subsection{Fixed points from fusion coefficients}
It is easy to understand that in general the constraint (\ref{fixed-poinfs-of-integer-spin-simple-current}) is a necessary but non sufficient condition for the fixed points. Hence the solutions found earlier might actually be too many, in the sense that some of them might not be fixed points. Then it is more useful to be a little bit more systematic and study the structure of the fixed points directly from the fusion coefficients.

\subsubsection{Twisted sector}
Let us start again from the twisted sector. For twisted fixed points we have to demand that
\begin{equation}
 N_{(J,\phi)\widehat{(f,\psi)}}^{\phantom{(J,\phi)\widehat{(f,\psi)}}\widehat{(f,\psi)}}=1\,.
\end{equation}
On the other hand, if $N$ is an arbitrary field of the orbifold theory, in terms of the $S$ and $P$ matrix of the original theory we have
\begin{eqnarray}
 N_{(J,\phi)\widehat{(f,\psi)}}^{\phantom{(J,\phi)\widehat{(f,\psi)}}\widehat{(f,\psi)}} 
 &=&  \sum_{N} 
 \frac{S_{(J,\phi)N}S_{\widehat{(f,\psi)}N}S_{\phantom{\dagger}N}^{\dagger\phantom{N}\widehat{(f,\psi)}}}{S_{(0,0)N}}=
 \nonumber\\  &=&  
 \sum_{(p,q)}
 \frac{S_{(J,\phi)(p,q)}S_{\widehat{(f,\psi)}(p,q)}S_{\phantom{\dagger}(p,q)}^{\dagger\phantom{(p,q)}\widehat{(f,\psi)}}}{S_{(0,0)(p,q)}}+
 \nonumber\\ &+&  
 \sum_{(j,\chi)}
 \frac{S_{(J,\phi)(j,\chi)}S_{\widehat{(f,\psi)}(j,\chi)}S_{\phantom{\dagger}(j,\chi)}^{\dagger\phantom{(j,\chi)}\widehat{(f,\psi)}}}{S_{(0,0)(j,\chi)}}+
 \nonumber\\ &+& 
 \sum_{\widehat{(p,\chi)}}
 \frac{S_{(J,\phi)\widehat{(p,\chi)}}S_{\widehat{(f,\psi)}\widehat{(p,\chi)}}S_{\phantom{\dagger}\widehat{(p,\chi)}}^{\dagger\phantom{\widehat{(p,\chi)}}\widehat{(f,\psi)}}}{S_{(0,0)\widehat{(p,\chi)}}}=
 \nonumber\\ &=& (BHS)= \nonumber\\ &=& 
 \frac{1}{2} \sum_j 
 \left[ \frac{(S_{Jj})2 }{(S_{0j})2} S_{fj} S_{\phantom{\dagger}j}^{\dagger\phantom{j}f}+
 e^{i\pi\phi} \frac{S_{Jj}P_{fj} P_{\phantom{\dagger}j}^{\dagger\phantom{j}f}}{S_{0j}}  \right]\,.
\end{eqnarray}
More in general one has
\begin{equation}
 N_{(J,\phi)\widehat{(f,\psi)}}^{\phantom{(J,\phi)\widehat{(f,\psi)}}\widehat{(f',\psi')}} =
 \frac{1}{2} \sum_j 
 \left[ \frac{(S_{Jj})2 }{(S_{0j})2} S_{fj} S_{\phantom{\dagger}j}^{\dagger\phantom{j}f'}+
 e^{i\pi(\phi+\psi-\psi')} \frac{S_{Jj}P_{fj} P_{\phantom{\dagger}j}^{\dagger\phantom{j}f'}}{S_{0j}}  \right]\,.
\end{equation}

It is important to remember that here we want $\widehat{(f,\psi)}$ to be a fixed point of $(J,\phi)$, i.e.
\begin{equation}
 N_{(J,\phi)\widehat{(f,\psi)}}^{\phantom{(J,\phi)\widehat{(f,\psi)}}\widehat{(f',\psi')}} = 
 \delta_f^{f'} \delta_\psi^{\psi'}\,.
\end{equation}
By itself, $f$ is not a fixed point of $J$ in the original theory.

Now use formula (\ref{simple-current-fundamental-relation}) in the first sum. In the following, we will restrict ourselves to order-2 simple currents. Because of the square and the fact that the monodromy charge of $j$ is either integer of half-integer\footnote{For order-2 simple currents.}, the exponent cancels out. Then we are left with $S$ times $S^\dagger$, which gives $\delta_f^{f'}$. 

We need to be more careful with the second piece, which involves the integer-valued \cite{Pradisi:1995qy,Gannon:1999is} $Y_{Jf}^{\phantom{Jf}f'}$-tensor. Our constraint reads then
\begin{equation}
 \delta_f^{f'} \delta_\psi^{\psi'} = \frac{1}{2}\delta_f^{f'} + 
 e^{i\pi(\phi+\psi-\psi')}\frac{1}{2}Y_{Jf}^{\phantom{Jf}f'}\,,
\end{equation}
which reduces to
\begin{equation}
 e^{i\pi\phi}Y_{Jf}^{\phantom{Jf}f'}=\delta_f^{f'}\qquad (\psi=\psi')
\end{equation}
or
\begin{equation}
 e^{i\pi(\phi+\psi-\psi')}Y_{Jf}^{\phantom{Jf}f'}=-\delta_f^{f'}\qquad (\psi\neq\psi')\,.
\end{equation}
Since we are considering currents with order 2, we can simplify the minus sign on the r.h.s. with $e^{i\pi(\psi-\psi')}$ on the l.h.s., thus re-obtaining the same expression of the case $\psi=\psi'$ for our constraint, which explicitly reads:
\begin{equation}
 e^{i\pi\phi} \sum_j \frac{S_{Jj}P_{fj} P_{\phantom{\dagger}j}^{\dagger\phantom{j}f'}}{S_{0j}}=\delta_f^{f'}\,.
\end{equation}
In order to solve it, let us study for the moment the equation:
\begin{equation}
 \sum_j x_j P_{fj} P_{\phantom{\dagger}j}^{\dagger\phantom{j}f'}=\delta_f^{f'}\,,
\end{equation}
for some $x_j$. Define a vector $v_f$ with components
\begin{equation}
 (v_f)_j:=x_j P_{fj}\,.
\end{equation}
Then we have
\begin{equation}
 \sum_j (v_f)_j P_{\phantom{\dagger}j}^{\dagger\phantom{j}f'}=\delta_f^{f'}\,.
\end{equation}
The vector $v_f$ is then orthogonal to all the columns of the matrix $P$, except for the column $f$ with which it has unit scalar product. Since $P$ is unitary, this implies that
\begin{equation}
 (v_f)_j=P_{fj}\,,
\end{equation}
which by definition yields\footnote{A shorter derivation is the following. Consider a diagonal matrix $X$ whose diagonal entries are $x_j$. Then the constraint in matrix form is: $PXP^\dagger=1$. Recalling that $PP^\dagger=1$ by unitarity, one can write $P(X-1)P^\dagger=0$, which gives the solution $X=1$.}
\begin{equation}
 x_j=1 \qquad \forall j\,.
\end{equation}

Going back to our situation where $x_j=e^{i\pi\phi} S_{Jj}/S_{0j}$, we arrive at the final form of our constraint:
\begin{equation}
 e^{i\pi\phi}S_{Jj}=S_{0j}\,.
\end{equation}
This is precisely the constraint (\ref{orbifold-twisted-fixed-poinfs-of-integer-spin-simple-current}) used previously. Hence the twisted fixed points are the same as before. This also shows that the previous constraint is a necessary and sufficient condition for the twisted fixed points of integer spin simple currents.

\subsubsection{Off-diagonal fields}
Similar arguments apply for the untwisted sector. Starting with off-diagonal fixed points one has
\begin{eqnarray}
 N_{(J,\phi)(p,q)}^{\phantom{(J,\phi)(p,q)}(p,q)} 
 &=& \sum_N 
 \frac{S_{(J,\phi)N}S_{(p,q)N}S_{\phantom{\dagger}N}^{\dagger\phantom{N}(p,q)}}{S_{(0,0)N}}=
 \nonumber\\  &=& 
 \sum_{(i,j)}
 \frac{S_{(J,\phi)(i,j)}S_{(p,q)(i,j)}S_{\phantom{\dagger}(i,j)}^{\dagger\phantom{(i,j)}(p,q)}}{S_{(0,0)(i,j)}}+
 \nonumber\\ &+&  
 \sum_{(i,\psi)}
 \frac{S_{(J,\phi)(i,\psi)}S_{(p,q)(i,\psi)}S_{\phantom{\dagger}(i,\psi)}^{\dagger\phantom{(i,\psi)}(p,q)}}{S_{(0,0)(i,\psi)}}+
 \nonumber\\ &+&  
 \sum_{\widehat{(i,\psi)}}
 \frac{S_{(J,\phi)\widehat{(i,\psi)}}S_{(p,q)\widehat{(i,\psi)}}S_{\phantom{\dagger}\widehat{(i,\psi)}}^{\dagger\phantom{\widehat{(i,\psi)}}(p,q)}}{S_{(0,0)\widehat{(i,\psi)}}}=
 \nonumber\\ &=&   (BHS) =\nonumber\\ &=&
 N_{Jp}^{\phantom{Jp}p} N_{Jq}^{\phantom{Jq}q} + N_{Jp}^{\phantom{Jp}q} N_{Jq}^{\phantom{Jq}p}\,.
\end{eqnarray}
This must be equal to $1$. Moreover $N_{ij}^{\phantom{ij}k}$ are positive integers. Hence we have two possibilities:
\begin{itemize}
 \item either
 \begin{equation}
\left\{
 \begin{array}{l}
   N_{Jp}^{\phantom{Jp}p} = N_{Jq}^{\phantom{Jq}q} = 1 \qquad \Rightarrow \,\,p \,\,\& \,\,q {\rm \,\,are\,\, fixed \,\,points\,\, of\,\,} J\\ 
   N_{Jp}^{\phantom{Jp}q} = N_{Jq}^{\phantom{Jq}p} = 0 
 \end{array}
\right.
\end{equation}
\item or
 \begin{equation}
 \left\{
  \begin{array}{l}
   N_{Jp}^{\phantom{Jp}p} = N_{Jq}^{\phantom{Jq}q} = 0 \\ 
   N_{Jp}^{\phantom{Jp}q} = N_{Jq}^{\phantom{Jq}p} = 1 \qquad \Rightarrow \,\,p \,\,\& \,\,q {\rm \,\,are\,\, in\,\,the\,\,same\,\,} J{\rm -orbit, i.e. \,\,} p=Jq
  \end{array}
 \right.
 \end{equation}
\end{itemize}
These two options are again the ones derived already before using the naive approach of the previous section (see discussion around (\ref{orbifold-offdiagonal-fixed-poinfs-of-integer-spin-simple-current}).).

\subsubsection{Diagonal fields}
For diagonal fixed points one has
\begin{eqnarray}
 N_{(J,\phi)(i,\psi)}^{\phantom{(J,\phi)(i,\psi)}(i,\psi)} 
 &=& \sum_N 
 \frac{S_{(J,\phi)N}S_{(i,\psi)N}S_{\phantom{\dagger}N}^{\dagger\phantom{N}(i,\psi)}}{S_{(0,0)N}}=
 \nonumber\\  &=& 
 \sum_{(p,q)}
 \frac{S_{(J,\phi)(p,q)}S_{(i,\psi)(p,q)}S_{\phantom{\dagger}(p,q)}^{\dagger\phantom{(i,\psi)}(i,\psi)}}{S_{(0,0)(p,q)}}+
 \nonumber\\ &+&  
 \sum_{(j,\chi)}
 \frac{S_{(J,\phi)(j,\chi)}S_{(i,\psi)(j,\chi)}S_{\phantom{\dagger}(j,\chi)}^{\dagger\phantom{(j,\chi)}(i,\psi)}}{S_{(0,0)(j,\chi)}}+
 \nonumber\\ &+&  
 \sum_{\widehat{(j,\chi)}}
 \frac{S_{(J,\phi)\widehat{(j,\chi)}}S_{(i,\psi)\widehat{(j,\chi)}}S_{\phantom{\dagger}\widehat{(j,\chi)}}^{\dagger\phantom{\widehat{(j,\chi)}}(i,\psi)}}{S_{(0,0)\widehat{(j,\chi)}}}=
 \nonumber\\ &=&   (BHS) =\nonumber\\ &=&
 \frac{1}{2} N_{Ji}^{\phantom{Ji}i}(N_{Ji}^{\phantom{Ji}i} + e^{i\pi\phi})\,.
\end{eqnarray}
Again we must demand 
\begin{equation}
 N_{(J,\phi)(i,\psi)}^{\phantom{(J,\phi)(i,\psi)}(i,\psi)} =1\,;
\end{equation}
then the only solution is when\footnote{We can exclude the other possibility $\phi=1$ and $N_{Ji}^{\phantom{Ji}i}=2$, because $J$ is a simple current.} $N_{Ji}^{\phantom{Ji}i}=1$, i.e. $i$ is a fixed point of $J$, and $\phi=0$, i.e. these fixed points appear only for the symmetric diagonal representation of the simple current. In the case of diagonal fields then, only a subset of the naive guess as in the previous section gives the correct fixed points.

\section{Example: $SU(2)_k$}
\label{Example SU(2)k}
Here we consider some examples of the previous general theory. We take our CFT to be an $SU(2)_k$ WZW model and work out spectrum and fusion rules of the orbifold theory.

Let us recall a few facts about affine Lie algebras \cite{Knizhnik:1984nr,Gepner:1986wi}. In an affine Lie algebra with group $G$, the weights of the highest weight representations $\lambda$ are given by
\begin{equation}
h(\lambda)=\frac{\frac{1}{2}C(\lambda)}{k+g}\,,
\end{equation}
where $C(\lambda)$ denotes the quadratic Casimir eigenvalue, $g$ is the dual Coxeter number (equal to half the Casimir of the adjoint representation) and $k$ is the level. The central charge is
\begin{equation}
c(G,k)=\frac{k\,{\rm dim}\,G}{k+g}
\end{equation}
and the matrix element is
\begin{equation}
S(\lambda,\mu)=const \cdot \sum_w \epsilon(w)\exp\left(-\frac{2\pi i}{k+g}(w(\lambda+\delta),\mu+\delta)\right)\,.
\end{equation}
Here the sum is over all the elements of the Weyl group and $\epsilon$ is the determinant of $w$. The normalization constant is fixed by unitarity and the requirement $S_{00}>0$.

Now we can apply these general pieces of information to our $SU(2)_k$ models (and later to $B(n)_1$ and $D(n)_1$ series).

\subsection{Generalities about $SU(2)_k$ WZW model}
In the $SU(2)_k$ theory, the level $k$ specifies both the central charge
\begin{equation}
 c=\frac{3 k}{k+2}
\end{equation}
and the spectrum of the primary fields through their weights
\begin{equation}
 h_{2j} =\frac{j (j+1)}{k+2},\qquad 2j=0,1,\dots k.
\end{equation}
Moreover, the field corresponding to the last value $2j=k$ is a simple current\footnote{Note that $j$ is either integer or half-integer.} of order two, the fusion being:
\begin{equation}
 (k)\times (2j) = (k-2j).
\end{equation}
Its weight is $h_{2j=k}=\frac{k}{4}$. This is integer or half-integer if $k$ is even. Furthermore, in the latter case, there is also a fixed point, given by the median value $2j=\frac{k}{2}$:
\begin{equation}
 (k)\times (\frac{k}{2}) = (\frac{k}{2}).
\end{equation}
There are no fixed points for odd $k$.

We can label these $k+1$ fields using their value of $j$. It will be convenient to call them
\begin{equation}
 \{\phi_{2j}\}=\{\phi_0 \equiv \mathbb{I} \,,\quad \phi_1\,,\quad\dots\,,\quad \phi_k\,\}. 
\end{equation}
The $S$ matrix is given by \cite{Cappelli:1986hf}
\begin{equation}
 S_{2j,\,2m}=\sqrt{\frac{2}{k+2}}\sin{\left[\frac{\pi}{k+2}(2j+1)(2m+1)\right]}.
\end{equation}

\subsection{$SU(2)_k \otimes SU(2)_k /\mathbb{Z}_2$ Orbifold: particular level}
Now let us consider the orbifold theory at some particular level $k$. The notation we will be using is as follows. First of all we need to distinguish the three types of fields in the orbifold theory: diagonal, off-diagonal and twisted fields.

Diagonal fields are generated by taking the symmetric tensor product of each field in the original theory with itself or the antisymmetric tensor product with the same field with its first non-vanishing descendant. Hence there are $2(k+1)$ diagonal fields, that will be denoted as:
\begin{equation}
 (2j,\psi)\qquad\psi=0,\,1
\end{equation}
with $2j=0,1,\dots k$. Here $\psi=0$ ($\psi=1$) for the symmetric (anti-symmetric) representation.
These fields have weights
\begin{equation}
 h_{(2j,\psi)}=2\,\frac{j(j+1)}{k+2}+\delta_{2j,0}\delta_{\psi,1}.
\end{equation}
The factor $2$ in front comes from the sum of weights of the fields appearing in the tensor product. In the anti-symmetric representation ($\psi=1$) of the identity ($2j=0$), one has to include the contribution to the weight coming from the Virasoro operators $L_{-1}$. The ground state is degenerate with dimension three due to the three $SU(2)$ generators.

Off-diagonal fields are obtained by taking the symmetric tensor product of each field in the original theory with a different field. Hence there are $\frac{k(k+1)}{2}$ non-diagonal fields, that will be denoted as:
\begin{equation}
 (\phi_{2i},\phi_{2j})\qquad 2i<2j.
\end{equation}
These fields have weights
\begin{equation}
 h_{(\phi_{2i},\,\phi_{2j})}=\frac{i(i+1)}{k+2}+\frac{j(j+1)}{k+2},
\end{equation}
which is simply the sum of the weights of the fields in the tensor product.

Twisted fields of any orbifold theory were described in \cite{Klemm:1990df}. After adapting their result (\ref{KS weight for twisted fields}) to our $\mathbb{Z}_2$ orbifold, we find that there are two twisted fields associated to each primary of the original theory. Hence there are $2(k+1)$ twisted fields, that will be denoted as:
\begin{equation}
 \widehat{(2j,m)}\qquad m=0,\,1,
\end{equation}
with $2j=0,1,\dots k$ as usual. Their weights are given by:
\begin{equation}
 h_{\widehat{(2j,m)}}=\frac{1}{2}\left[\frac{j(j+1)}{k+2}+m\right]+\frac{3k}{16(k+2)}.
\end{equation}

The next step is to compute the $S$ matrix for this orbifold theory using the BHS formulas (\ref{BHS off-diag}, \ref{BHS diag}, \ref{BHS twisted}). Using the Verlinde formula \cite{Verlinde:1988sn} we will then be able to compute the fusion rules, which will allow us to look for simple currents in the orbifold theory. In appendix \ref{Tables for SU(2)k Orbifold} we summarize the simple currents and corresponding fixed points for particular values of the level $k$. We will consider only ``integer spin'' simple currents, namely those with integer weight.

\subsection{$SU(2)_k \otimes SU(2)_k /\mathbb{Z}_2$ Orbifold: generic level}
\label{su2 fixed point table}
From the results corresponding to a few values of $k$, we can determine important generalizations for arbitrary $k$.\\
First of all, for all $k$ there is at least one non-trivial integer spin simple current, namely $(0,1)$ with $h=1$, whose fixed points are all the off-diagonal fields. Their number is $\binom{k+1}{2}=\frac{k(k+1)}{2}$.\\
In addition, if $k$ is even, there are  other two integer spin simple currents\footnote{These are actually the only ones with integer spin.}. They are the symmetric and anti-symmetric diagonal fields corresponding to the last value $2j=k$: $(k,0)$ and $(k,1)$, both with $h=\frac{k}{2}$. This reflects the general structure of the $SU(2)_k$ simple currents. Their fixed points are also easily determined. For the current $(k,0)$ they come from diagonal, off-diagonal and twisted fields according to some rules which are given below, while those of $(k,1)$ come only from off-diagonal and twisted fields.

Summarizing:

\begin{tabular}{l l}
&\\
 Simple current & Fixed point\\
\hline
$(0,1)$, \, $h=1$ & \textit{all the $\frac{k(k+1)}{2}$ off-diagonal fields}\\
$(k,0)$, \, $h=\frac{k}{2}$ & \textit{$2$ diag. + $\frac{k}{2}$ off-diag. + $(k+2)$ twisted fields}\\
$(k,1)$, \, $h=\frac{k}{2}$ & \textit{$\frac{k}{2}$ off-diag. + $k$ twisted fields}\\
&\\
\end{tabular}

The rule to construct the fixed points of the additional simple currents when $k$ is even is as follows. 

The diagonal fields appearing as fixed points of $(k,0)$ are always the two fields in the middle: $(\frac{k}{2},0)$ and $(\frac{k}{2},1)$. These are $\frac{k}{2}$ and have weights
\begin{equation}
 h_{(\frac{k}{2},0)}=h_{(\frac{k}{2},1)}=\frac{1}{8}\frac{k(k+4)}{k+2}\,.
\end{equation}

The off-diagonal fields appearing as fixed points are the same for both the two additional currents and are given by the fields $(\phi_{2i} ,\phi_{k-2i})$, i.e. the fields $2i$ and $k-2i$ belong to the same orbit under $J\equiv \phi_k$. The weights of these off-diagonal fixed points are:
\begin{equation}
 h_{(\phi_{2i} ,\phi_{k-2i})}=\frac{1}{k+2}\left[i^2+\left(\frac{k}{2}-i\right)^2+\frac{k}{2}\right],
\end{equation}
with $2i=0,1,\dots,k$.

The fixed points coming from the twisted sector are ``complementary'' for the two additional simple currents, in the sense that $(k,0)$ has $\widehat{(4j,m)}$, $m=0,\,1$ and $2j=0,1,\dots,k$, as fixed points\footnote{Explicitly, these fixed points are $\widehat{(0,m)},\,\,\widehat{(2,m)},\,\,\widehat{(4,m)},\,\dots,\,\widehat{(k,m)}$, $m=0,\,1$, with the first argument even. In total, $k+2$.}, while $(k,1)$ has $\widehat{(4j+1,m)}$, $m=0,\,1$ and $2j=0,1,\dots,k-1$, as fixed points\footnote{Explicitly, these fixed points are $\widehat{(1,m)},\,\,\widehat{(3,m)},\,\,\widehat{(5,m)},\,\dots,\,\widehat{(k-1,m)}$, $m=0,\,1$, with the first argument odd. In total, $k$.}. Their weights are:
\begin{equation}
 h_{\widehat{(4j,m)}}=\frac{1}{2}\left[\frac{2j(2j+1)}{k+2}+m\right]+\frac{3}{16(k+2)}
\end{equation}
and
\begin{equation}
 h_{\widehat{(4j+1,m)}}=\frac{1}{2}\left[\frac{1}{k+2}\left(2j+\frac{1}{2}\right)\left(2j+\frac{1}{2}+1\right)+m\right]+\frac{3}{16(k+2)}
\end{equation}
for $\widehat{(4j,m)}$ and $\widehat{(4j+1,m)}$ respectively.

As last remark, let us stress that all this structure agrees with the general theory of the previous section.

\section{Fixed point resolution in $SU(2)_k$ orbifolds}
\label{Fixed point resolution in SU(2)k orbifolds}
We would like to determine the $S^J$ matrices corresponding to the simple currents given above using formula (\ref{main formula for f.p. resolution}) which relates them to the $S$ matrix of the extended theory via the group characters $\Psi_i(J)$. As we will now explain, we know what the $S^J$ matrix is in the case $J\equiv (0,1)$. It is given by an expression analogous to the off-diagonal/off-diagonal BHS $S$ matrix, but with a minus (instead of the plus) sign. This is a fortunate situation because the current $J\equiv (0,1)$ is omnipresent, since it appears for all values of the level $k$. The other two currents that appear occasionally are slightly more complicated since they involve twisted fields.

In reading this section, the reader might find it useful to consult appendix \ref{appendix SU2 orbifolds}.

\subsection{$S^J$ matrices}
\subsubsection{$S^J$ matrix for $J\equiv (0,1)$}
The general procedure when we make an extension via integer spin simple currents is as follows: keep states that are invariant under the symmetry generated by the current, namely those with integer monodromy charge w.r.t. $J$, and organize fields into orbits. Fixed points are particular orbits: orbits with length one. 

Consider the current $J\equiv (0,1)$ of order $2$. The extension projects out the twisted fields, since they are all non-local w.r.t. this current. Only untwisted fields are left, both diagonal and off-diagonal. Off-diagonal fields are fixed points of $(0,1)$, so they get doubled by the extension, while diagonal fields group themselves into orbits of length two containing symmetric and anti-symmetric representation of each original field. It is interesting to see that the resulting theory is equal to the tensor product $SU(2)_k\otimes SU(2)_k$. What happens is the following. The length-two orbits come from diagonal fields and correspond to fields $\phi_{2i}\otimes \phi_{2i}$ of the tensor product, while the two fields coming from the fixed points correspond to $\phi_{2i}\otimes \phi_{2j}$ and $\phi_{2j}\otimes \phi_{2i}$ (with $2i \neq 2j$) of the tensor product. The weights indeed match exactly. So in the end we have the result:
\begin{equation}
\left(\mathcal{A}\otimes \mathcal{A}/\mathbb{Z}_2 \right)_{(0,1)}=\mathcal{A}\otimes \mathcal{A} ·
\end{equation}
The subscript $(0,1)$ means that we are taking the extension by the $(0,1)$ current.
This result is not limited to $\mathcal{A}=SU(2)_k$, but is true for any rational CFT. The reason is that this
simple current extension is in fact the inverse of the permutation orbifold procedure. This follows from the
fact that the permutation orbifold splits the original chiral algebra in a symmetric and an anti-symmetric part,
and the representation space of the current $(0,1)$ is precisely the latter. By extending the chiral algebra with
this current we re-constitute the original chiral algebra of $\mathcal{A}\otimes \mathcal{A}$. This result extends
straightforwardly to the other representations, and of course the twisted field must be projected out, since by construction 
they are
non-local with respect to $\mathcal{A}\otimes \mathcal{A}$.

Resolving the fixed points is equivalent to finding a set of $S^J$ matrices such that
\begin{equation}
\label{main formula for f.p. resolution 2}
\tilde{S}_{(a,i)(b,j)}=\frac{|G|}{\sqrt{|U_a||S_a||U_b||S_b|}}\sum_{J\in G}\Psi_i(J) S^J_{ab} \Psi_j(J)^{\star}\,,
\end{equation}
where $\tilde{S}$ is the full extended $S$ matrix, $a$ and $b$ denote the fixed points of $J$, while $i$ and $j$ the fields into which the fixed points are resolved. For $J\equiv (0,1)$ we know that the extended theory is the tensor product theory, whose $S$ matrix is the tensor product of the $S$ matrices of the two factors. When we extend w.r.t. $(0,1)$ only two terms contribute on the r.h.s., namely $S^0\equiv S^{BHS}$ and $S^J$. The indices $a$ and $b$ run over the off-diagonal fields. Hence it is natural to write down the following ansatz for $S^J$ for $J=(0,1)$:
\begin{equation}
\label{SJ_(offdiag.-offdiag.)}
\boxed{S^J_{(mn)(pq)}=S_{mp}S_{nq}-S_{mq}S_{np}}\,.
\end{equation}
This is unitary and satisfies the modular constraint $(S^JT^J)^3=(S^J)^2$. Here $S_{mp}$ is the $S$ matrix of the original theory\footnote{As an exercise, one could try to write this $S^J$ matrix explicitly for $k=2$. With our conventional choice for the labels of the fields, it turns out to be numerically equal to minus the $S$ matrix of the original $SU(2)_2$ theory isomorphic to the Ising model: $S^J=-S_{SU(2)_2}$.\label{footnote_SJ}}. Note that there is an apparent sign ambiguity: the matrix elements
depend on the labelling of the off-diagonal fields, because the field $(p,q)$ might just as well have been labelled $(q,p)$. 
This is irrelevant, since it merely amounts to a basis choice among the two split fields originating from $(q,p)$. It is
easy to check that the matrix $\tilde S$ computed with  (\ref{main formula for f.p. resolution}) is indeed the one
of the tensor product, {\it i.e.} $S_{mp}S_{nq}$.

\subsubsection{$S^J$ matrix for $J\equiv (k,0)$}
The order-$2$ current $J\equiv (k,0)$ arises only when $k$ is even, so in this subsection we will restrict to such values. The first thing we need to do is to determine the orbits of the current, since they become the fields of the extended theory.

Either by looking at explicit low values of $k$ or by general arguments, one can observe a few facts about orbits of $J\equiv (k,0)$.\\
First, form the diagonal sector, $J$ couples symmetric (anti-symmetric) representation of a field $\phi_{2j}$ with symmetric (anti-symmetric) representation of its image $J\cdot \phi_{2j}=\phi_{k-2j}$ into length-2 orbits. In particular, the field $(\frac{k}{2},0)$ can couple only to itself, hence it must be a fixed point. Similarly for the field $(\frac{k}{2},1)$. So, there are exactly $k$ length-2 orbits and two fixed points coming from diagonal fields.\\
Secondly, from the off-diagonal sector, only $(\phi_{2i},\phi_{2j})$ with $2i$ and $2j$ either both even or both odd  survive the projection, because only those have a well-defined monodromy charge. Moreover, $J$ couples the field $(\phi_{2i},\phi_{2j})$ with its image $J\cdot(\phi_{2i},\phi_{2j})=(\phi_{k-2i},\phi_{k-2j})$. In particular, fields of the form $(\phi_{2j},\phi_{k-2j})$ must be fixed points. There are $\frac{1}{2}\left((\frac{k}{2})^2-\frac{k}{2}\right)$ length-2 orbits and $\frac{k}{2}$ fixed points coming from off-diagonal fields. In this formula, we divide by $2$ because generically fields are coupled into orbits. The contribution within brackets comes from the number of off-diagonal fields that are not projected out minus the number of off-diagonal fixed points.\\
Finally, there are no orbits coming from the twisted sector, but only $k+2$ fixed points.

Putting everything together, the theory extended by $J\equiv (k,0)$ has $3k+8$ fixed points (i.e. twice the number given in section \ref{su2 fixed point table}) plus $\frac{k(k+6)}{8}$ length-2 orbits.

Here an ansatz for $S^J$ is still unknown for generic values of the level $k$. We have so far worked out only the simpler case $k=2$, which is closely related to the Ising model.

\subsubsection{$S^J$ matrix for $J\equiv (k,1)$}
Also in this case $k$ must be even in order for the current $J\equiv (k,1)$ to be present. The orbit structure here is, \textit{mutatis mutanda}, analogous to the previous one.\\
From the diagonal sector, $J$ couples symmetric (anti-symmetric) representation of a field $\phi_{2j}$ with anti-symmetric (symmetric) representation of its image $J\cdot \phi_{2j}=\phi_{k-2j}$ into length-2 orbits. In particular, the fields $(\frac{k}{2},0)$ and $(\frac{k}{2},1)$ must couple to each other, contributing an additional orbit. There are exactly $k+1$ length-2 orbits and no fixed points coming from diagonal fields.\\
From the off-diagonal sector, one has the same length-2 orbits as for the previous case above. So there are again $\frac{1}{2}\left((\frac{k}{2})^2-\frac{k}{2}\right)$ orbits and $\frac{k}{2}$ fixed points coming from off-diagonal fields.\\
As above, there are no orbits coming from the twisted sector, but only $k$ fixed points.

Putting everything together, the theory extended by $J\equiv (k,1)$ has $3k$ fixed points (i.e. twice the number as given in section \ref{su2 fixed point table}) plus $\frac{k(k+6)}{8}+1$ length-2 orbits.

Also here an ansatz for $S^J$ is unknown, except for the case $k=2$, given below.

\subsection{$S^J$ matrices for $k=2$}
\label{SJ of SU2 level 2}
The case $k=2$ is particularly simple to analyze, because the matrices involved are relatively small, but it is also very interesting, because it gives us a lot of insights. The orbit structure of this permutation orbifold is given in the appendix.

First of all, as we have already remarked in footnote \ref{footnote_SJ},
\begin{equation}
S^{J\equiv (0,1)}=-S_{SU(2)_2}\,,
\end{equation}
resolving the three fixed points of the current $(0,1)$ (see table \ref{table S^J1_k=2}). It is important to remark here that the form of the $S^J$ matrix depends very much on the choice of the labels for the mother CFT: once we reshuffle the labeling of the original $SU(2)_2$ spectrum, the $S^J$ does not simply change by a reshuffling of its rows and columns since some entries can drastically change as well.
\begin{table}[ht]
\caption{Fixed point Resolution: Matrix $S^{J\equiv (0,1)}$}
\centering
\begin{tabular}{c|c c c}
\hline \hline\\
$S^{J\equiv (0,1)}$ & $(\phi_0,\phi_1)$ & $(\phi_0,\phi_2)$ & $(\phi_1,\phi_2)$ \\ 
\hline &&&\\
$(\phi_0,\phi_1)$   & $-\frac{1}{2}$        & $-\frac{\sqrt{2}}{2}$  & $-\frac{1}{2}$         \\
$(\phi_0,\phi_2)$   & $-\frac{\sqrt{2}}{2}$ & $0$                    & $\frac{\sqrt{2}}{2}$   \\
$(\phi_1,\phi_2)$   & $-\frac{1}{2}$        & $\frac{\sqrt{2}}{2}$   & $-\frac{1}{2}$
\end{tabular}
\label{table S^J1_k=2}
\end{table}

By numerical checks of unitarity and modular properties\footnote{Namely, one checks that $S^J$ satisfies $S^J (S^J)^\dagger=1$ and $(S^J T^J)^3=(S^J)^2$.}, one can also guess the $S^J$ matrix of the third current $(2,1)$:
\begin{equation}
S^{J\equiv (2,1)}=-S_{SU(2)_2}\,.
\end{equation}
This is numerically equal to the previous one if we order the fixed point fields according to their conformal weights  in the same way as for the first current (see table \ref{table S^J3_k=2}). Indeed, the origin of this equality is that these two extensions are isomorphic to each other, having their fixed points and orbits equal weights.
\begin{table}[ht]
\caption{Fixed point Resolution: Matrix $S^{J\equiv (2,1)}$}
\centering
\begin{tabular}{c|c c c}
\hline \hline\\
$S^{J\equiv (2,1)}$ & $\widehat{(1,0)}$ & $(\phi_0,\phi_2)$ & $\widehat{(1,1)}$ \\ 
\hline &&&\\
$\widehat{(1,0)}$   & $-\frac{1}{2}$        & $-\frac{\sqrt{2}}{2}$  & $-\frac{1}{2}$         \\
$(\phi_0,\phi_2)$   & $-\frac{\sqrt{2}}{2}$ & $0$                    & $\frac{\sqrt{2}}{2}$   \\
$\widehat{(1,1)}$   & $-\frac{1}{2}$        & $\frac{\sqrt{2}}{2}$   & $-\frac{1}{2}$
\end{tabular}
\label{table S^J3_k=2}
\end{table}

To determine the $S^J$ matrix of the second current $(2,0)$ is a bit more complicated. We would like to use the main formula (\ref{main formula for f.p. resolution 2}) where we need the $S$ matrix of the extended theory. Observe that the extended theory has 16 primaries, of which $2 \times 7$ come from the seven fixed points of $J$, all with known conformal weights. Moreover, it also has central charge $c\leq 3$. There are not many options one has to consider. Indeed, one can show that the extended theory coincides with the tensor product theory $SU(3)_1 \times U(1)_{48}$ extended with a particular integer spin simple current of order three. We denote it here by $(1,16)$. It has no fixed points and its $S$ matrix is known. Using (\ref{main formula for f.p. resolution 2}), we can now determine the unknown $S^{J\equiv (2,0)}$ by brute-force calculation. The result is given in table \ref{table S^J2_k=2} (more details in appendix \ref{appendix SU2 orbifolds}).
\begin{table}[ht]
\caption{Fixed point Resolution: Matrix $S^{J\equiv (2,0)}$}
\centering
\begin{tabular}{c|c c c c c c c}
\hline \hline\\
$S^{J\equiv (2,0)}$ & $(1,0)$ & $(1,1)$ & $(\phi_0,\phi_2)$ & $\widehat{(0,0)}$ & $\widehat{(0,1)}$ & $\widehat{(2,0)}$ & $\widehat{(2,1)}$ \\ 
\hline &&&\\
$(1,0)$             & $2 i a$  & $2 i a$  & 0        & $2 i b$  & $-2 i b$ & $-2 i b$ & $2 i b$ \\
$(1,1)$             & $2 i a$  & $2 i a$  & 0        & $-2 i b$ & $2 i b$  & $2 i b$  & $-2 i b$\\
$(\phi_0,\phi_2)$   & 0        & 0        & 0        & $2 i a$  & $-2 i a$ & $2 i a$  & $-2 i a$\\
$\widehat{(0,0)}$   & $2 i b$  & $-2 i b$ & $2 i a$  & $-2 i d$ & $-2 i d$ & $2 i c$  & $2 i c$ \\
$\widehat{(0,1)}$   & $-2 i b$ & $2 i b$  & $-2 i a$ & $-2 i d$ & $-2 i d$ & $2 i c$  & $2 i c$\\
$\widehat{(2,0)}$   & $-2 i b$ & $2 i b$  & $2 i a$  & $2 i c$  & $2 i c$  & $2 i d$  & $2 i d$\\
$\widehat{(2,1)}$   & $2 i b$  & $-2 i b$ & $-2 i a$ & $2 i c$  & $2 i c$  & $2 i d$  & $2 i d$
\end{tabular}
\label{table S^J2_k=2}
\end{table}
The numbers $a,\,b,\,c,\,d$ above are given by: $a=\frac{1}{4}$, $b=\frac{1}{4\sqrt{2}}$, $c=\frac{\sqrt{2-\sqrt{2}}}{8}$, $d=\frac{\sqrt{2+\sqrt{2}}}{8}$. One can check that the matrix above is unitary, modular invariant and produces sensible fusion coefficients.

A few remarks are in order. First, it is interesting to observe that the numbers $a$ and $b$ are related to the $S$ matrix of the original $SU(2)_2$ CFT, while $c$ and $d$ come from the corresponding $P$ matrix, $P=T^{1/2}ST^2ST^{1/2}$.

Second, as it was derived, this matrix is not the only possible one. There in fact exists a few other
consistent\footnote{I.e. unitary, modular invariant and with non-negative integer fusion coefficients.} possibilities for $S^J$ where some entries have different sign. The reason for this is the ambiguity existing in choosing which of the two splitted fixed points is number one and which number two. This is important since, due to the group characters appearing in (\ref{main formula for f.p. resolution 2}), a flip in their order would produce a sign flip in the corresponding entry of the $S^J$ matrix. So all the non-diagonal entries are determined up to this sign ambiguity.

\section{Fixed point resolution in $SO(N)_1$ orbifolds}
\label{Fixed point resolution in SO(N)1 orbifolds}
Another interesting example of fixed point resolution that we have worked out is the $SO(N)_1$ permutation orbifold. This is a relatively straightforward case since we know the extended theories of all of its integer spin simple current extensions. In fact, they can be derived from the same arguments given in section \ref{SJ of SU2 level 2} for the $SU(2)_2$ permutation orbifold. In the easier cases, the $S^J$ matrix can be computed using (\ref{SJ_(offdiag.-offdiag.)}), since the extension of the orbifold theory gives back the tensor product theory (or a theory isomorphic to it); in more complicated situations, the $S^J$ matrix can be derived from (\ref{main formula for f.p. resolution 2}) and the knowledge of the full, i.e. extended, $S$ matrix via the embedding that we have mentioned before. This embedding works as follows:
\begin{equation}
\label{SO embedded in SU}
\xymatrix{ 
 SO(N)_{perm} \ar[r] \ar[dr]_{ext} & SO(2N) \\
   & SU(N)\times U(1)  \ar[u]_{ext'}  }
\end{equation}
i.e. the extension of the permutation orbifold gives $SU(N)\times U(1)$ whose extension (with another particular current) is $SO(2N)$, the group where the permutation orbifold is embedded.

Let us remind a few facts about these two CFT's \cite{Knizhnik:1984nr,Gepner:1986wi}. The $U(1)_R$ CFT at radius $R$ has central charge $c=1$, $R$ primary fields labelled by $u=0,1,\dots,R-1$ with weight
\begin{equation}
h_u=\frac{u^2}{2R}\,\,{\rm mod}\, \mathbb{Z}.
\end{equation}
Its $S$ matrix and corresponding fusion rules are given by
\begin{eqnarray}
S_{uu'}=\frac{1}{\sqrt{R}}\,e^{-2\pi i \frac{uu'}{R}},\\
(u)\cdot(u')=(u+u')\,\,{\rm mod}\,R.
\end{eqnarray}
The $SU(N)_1=A(N-1)_1$ CFT has central charge $c=N-1$, $N$ primary fields labelled by $s=0,1,\dots,N-1$ with weight
\begin{equation}
h_s=\frac{s^2(N-1)}{2N}\,\,{\rm mod}\, \mathbb{Z}.
\end{equation}
Its $S$ matrix and corresponding fusion rules are given by
\begin{eqnarray}
S_{ss'}=\frac{1}{\sqrt{N}}\,e^{2\pi i \frac{ss'}{N}},\\
(s)\cdot(s')=(s+s')\,\,{\rm mod}\,N.
\end{eqnarray}

For our study of $SO(N)$ at level one, we only need to determine the level of the $SU(N)$ and the radius of the $U(1)$ factors. After a few trials, it is not difficult to convince ourselves that the level of the $SU(N)$ factor is one and the radius of the $U(1)$ factor is $16 N$, while the integer spin simple current (with order $N$) that we need to extend this product group in order to get $SO(2N)$ is\footnote{It is convenient to label fields in the tensor product by pairs $(s,u)$, with $s$ and $u$ labeling fields of the two factors. Sometimes other labels can be used, e.g. one single label $l$, with $l=s\cdot R + u$ or vice versa $s=l\,{\rm mod}\,R$ and $u=\left[\frac{l}{R}\right]$, squared brackets denoting the integer part.} $(\#,16)$, where the first entry denotes a particular field of the $SU(N)_1$ CFT depending\footnote{E.g. for low values of $N$, $\#=4$.} on the value $N$ and the second entry another particular, but given, field of the $U(1)_{16N}$ CFT. Explicitly,
\begin{equation}
(SO(N)_1\times SO(N)_1/\mathbb{Z}_2)_{\rm ext}=(SU(N)_1\times U(1)_{16N})_{\rm ext'}\,.
\end{equation}
The $S$ matrix of the tensor product theory is simply the tensor product of the two $S$ matrices, $S^{\otimes}_{(s,u)(s',u')}=S_{ss'}S_{uu'}$, while the $S$ matrix of the extended theory, $\tilde{S}$, is the tensor product $S$ matrix multiplied by the order $N$ of the current \cite{Schellekens:1990xy}. Hence the $S$ matrix of the extended tensor product $(SU(N)_1\times U(1)_{16N})_{(\#,16)}$ is:
\begin{equation}
\label{S matrix of ext SUN times U1}
\tilde{S}_{(su)(s'u')}=\frac{1}{4}\,\exp\left\{\frac{2\pi i}{N}\left(ss'-\frac{uu'}{16}\right)\right\}\,,
\end{equation}
where the factor $N$ in the denominator is cancelled by the order $N$ in the numerator. This gives the following fusion rules:
\begin{equation}
 (s,u)\cdot (s',u')=((s+s')\,{\rm mod}\,N,(u+u')\,{\rm mod}\,16N)\,.
\end{equation}

Recall that in the extended theory only certain fields $(s,u)$ appear, namely those with integer monodromy charge with respect to the current $(\#,16)$. It is given by 
\begin{equation}
\label{monodromy in SO(N)}
 Q_{(\#,16)}(s,u)=-\frac{\#\cdot s(N-1)+u}{N} \,{\rm mod}\,\mathbb{Z}\,.
\end{equation}
This allows us to analytically relate the labels $s$ and $u$ of the fields in the extension to the fields in the permutation orbifolds, by comparing the weights of the fields in the permutation, $\{h_{\rm perm}\}$, with the ones in the extension, $h_{s,u}=h_s+h_u$, and choosing $s$ and $u$ such that (\ref{monodromy in SO(N)}) is satisfied. This will be crucial when we use (\ref{main formula for f.p. resolution 2}).

Let us move now to study the fixed point resolution of the $SO(N)_1$ permutation orbifolds, distinguishing the case of $N$ even and $N$ odd.

\subsection{$B(n)_1$ series}
The $B(n)_1=SO(N)_1$, $N=2n+1$, series has central charge $c=\frac{N}{2}$ and three primary fields $\phi_i$ with weight $h_i=0,\frac{1}{2},\frac{N}{16}$ ($i=0,1,2$ respectively). The $S$ matrix is the same as the Ising model, as shown in table \ref{table S_Bn_1}.
\begin{table}[ht]
\caption{$S$ matrix for $B(n)_1$}
\centering
\begin{tabular}{c|c c c}
\hline \hline\\
$S_{B(n)_1}$ & $h=0$ & $h=\frac{1}{2}$ & $h=\frac{N}{16}$ \\ 
\hline &&&\\
$h=0$             & $\frac{1}{2}$        & $\frac{1}{2}$         & $\frac{\sqrt{2}}{2}$    \\
$h=\frac{1}{2}$   & $\frac{1}{2}$        & $\frac{1}{2}$         & $-\frac{\sqrt{2}}{2}$   \\
$h=\frac{N}{16}$  & $\frac{\sqrt{2}}{2}$ & $-\frac{\sqrt{2}}{2}$ & $0$
\end{tabular}
\label{table S_Bn_1}
\end{table}

The $B(n)_1$ series has two simple currents\footnote{And only two, because $N$ is odd. This will be different for the $D(n)_1$ series.}, namely the fields with $h_0=0$ (the identity) and $h_1=\frac{1}{2}$. 
In the tensor product they give rise to integer spin simple currents and can both be used to extend the permutation orbifold. Hence, according to our notation, $(B(n)_1)_{\rm perm}$ has four integer spin simple currents arising from the symmetric and anti-symmetric representations of $\phi_0$ and $\phi_1$. Explicitly they are: $(0,0)$, $(0,1)$, $(1,0)$ and $(1,1)$. This situation is very similar to the one already studied in section \ref{SJ of SU2 level 2}. The extension w.r.t. the identity $(0,0)$ is trivial. The extension w.r.t. the current $(0,1)$ projects out all the twisted sector and gives back the tensor product theory $B(n)_1\times B(n)_1$; the fixed points are all the three off-diagonal fields ($h_{01}=\frac{1}{2}$, $h_{02}=\frac{N}{16}$, $h_{12}=\frac{N}{16}+\frac{1}{2}$) and hence the corresponding $S^J$, with $J=(0,1)$, is given by (\ref{SJ_(offdiag.-offdiag.)}). \\
Also easy is the extension w.r.t. the current $(1,1)$: it is indeed isomorphic to the previous one. The fixed points are the off-diagonal field $(\phi_0,\phi_1)$ ($h=\frac{1}{2}$) and the two twisted fields coming from $\phi_2$ (with $h=\frac{N}{16}$ and $\frac{N}{16}+\frac{1}{2}$). All their weights are equal to the weights of the fixed points of the current $(0,1)$, hence,  if we label them according to $h$, the $S^J$ matrix for the  current $(1,1)$ is numerically the same as for $(0,1)$. \\
A bit more involved is the $S^J$ matrix for the current $(1,0)$. For this, we need to use the main formula (\ref{main formula for f.p. resolution 2}).

\subsubsection{$(B(n)_1)_{\rm perm}$ $S^J$ matrix for $J=(1,0)$}
There are seven fixed points for the current $J=(1,0)$ of the permutation orbifold $(B(n)_1)_{\rm perm}$, coming from all possible sectors. From the  diagonal fields, we have $(2,0)$ and $(2,1)$ (both have $h=\frac{N}{8}$), from the off-diagonal $(\phi_0,\phi_1)$ (with $h=\frac{1}{2}$) and from the twisted $\widehat{(0,0)}$ ($h=\frac{N}{32}$), $\widehat{(0,1)}$ ($h=\frac{N}{32}+\frac{1}{2}$), $\widehat{(1,0)}$ ($h=\frac{N+8}{32}$) and $\widehat{(1,1)}$ ($h=\frac{N+8}{32}+\frac{1}{2}$). We know the original $S$ matrix for these fields, given by $S^{BHS}$. We also know the $S$ matrix of the extended theory, $\tilde{S}$ as in (\ref{S matrix of ext SUN times U1}), given by the embedding (\ref{SO embedded in SU}). Hence we can use the simplified version (\ref{main formula for f.p. resolution simplified}) of the main formula (\ref{main formula for f.p. resolution 2}) as given in appendix \ref{appendix SU2 orbifolds} to obtain the desired matrix.

Before giving the $S^J$ matrix, there is a very important issue that we should cover first. We mentioned before that the labels of the permutation and those of the extension are different but related. How can we exactly relate them? Recall in the extension, fields are defined by orbits of the current, with all the fields in the same orbit having same weight (modulo integer) and same $S$ matrix (see \cite{Schellekens:1990xy}). Within each orbit in the extended theory, we choose the field with lowest weight as representative of the splitted fields coming from the fixed point resolution. According to this convention, every fixed point gets splitted in two fields $(s_1,u_1)$ and $(s_2,u_2)$ given by:
\begin{eqnarray}
\bullet & {\rm if \,\,n=3,4,7,8,11,12,\dots} & \,\Leftrightarrow\, 
{\rm if\,\,} \left[\frac{n-1}{2}\right] {\rm is \,\, odd \,\,} \nonumber\\
(2,0) \,\, \longrightarrow&  (0,2N) & \&\qquad (0,14N) \nonumber \\
(2,1) \,\, \longrightarrow&  (2,14N+8) & \&\qquad (N-2,2N-8) \nonumber \\
&&\nonumber\\
\bullet & {\rm if \,\,n=5,6,9,10,13,14\dots} & \,\Leftrightarrow\, 
{\rm if\,\,} \left[\frac{n-1}{2}\right] {\rm is \,\, even \,\,} \nonumber \\
(2,0) \,\, \longrightarrow&  (2,14N+8) & \&\qquad (N-2,2N-8) \nonumber \\
(2,1) \,\, \longrightarrow&  (0,2N) & \&\qquad (0,14N) \nonumber \\
&&\\
\bullet & {\rm for \,\, all \,\, n} & \nonumber \\
(\phi_0,\phi_1) \,\, \longrightarrow&  (1,4) & \&\qquad (N-1,16N-4) \nonumber \\
\widehat{(0,0)} \,\, \longrightarrow&  (0,N) & \&\qquad (0,15N) \nonumber \\
\widehat{(0,1)} \,\, \longrightarrow&  (2,15N+8) & \&\qquad (N-2,N-8) \nonumber \\
\widehat{(1,0)} \,\, \longrightarrow&  (N-1,N-4) & \&\qquad (1,15N+4) \nonumber \\
\widehat{(1,1)} \,\, \longrightarrow&  (3,12-N) & \&\qquad (1,N+4) \nonumber \\
&&\nonumber
\end{eqnarray}
This table also fixes the order of which field we call ``splitted field $1$'' and ``splitted field $2$''. We must use fields only from the first set or only from the second set when computing $S^J$. Both the two sets will give the same result, but we cannot choose field representative randomly without losing unitarity and/or modular invariance. It is interesting to check that the orbits corresponding to the two splitted fields are ``conjugate'' of each other, in the sense that $s_1+s_2 =0$ mod $N$ and $u_1+u_2=0$ mod $16 N$.

The $S^J$ matrix is now given below. It is expressed in terms of the $S$ and $P$ matrices\footnote{The $P$ matrix is $P=T^{1/2}ST^2ST^{1/2}$ and for the $B(n)_1$ series reads:
\begin{equation}
 P=\left(
\begin{array}{ccc}
\cos\left(\frac{\pi N}{8}\right) & \sin\left(\frac{\pi N}{8}\right)  & 0 \\
\sin\left(\frac{\pi N}{8}\right) & -\cos\left(\frac{\pi N}{8}\right) & 0 \\
0                                & 0                                 & 1 
\end{array}
\right)\,,\nonumber
\end{equation}
where $N=2n+1$.}
of the mother $B(n)_1$ theory; also a sign $\epsilon$ appears, depending on the value of $N=2n+1$, $\epsilon=(-1)^{\left[\frac{n-1}{2}\right]}$, square brackets denoting the integer part. We have checked that it is unitary ($S^J (S^J)^{\dagger}=1$), modular invariant ($(S^J T^J)^3=-1=(SJ)^2$) and it gives correct fusion coefficients.

\begin{eqnarray}
S^J_{(2,0)(2,0)} &=& -\frac{1}{2}\,i^N \nonumber \\
S^J_{(2,0)(2,1)} &=& \frac{1}{2}\,i^N \nonumber \\
S^J_{(2,0)(\phi_0,\phi_1)} &=& -\frac{1}{2} -S_{20}\,S_{21}
                        \,\,=\,\, 0 \nonumber \\
S^J_{(2,0)\widehat{(0,0)}} &=& -\epsilon \,\frac{1}{2} \,e^{\frac{\epsilon\pi iN}{4}} -\frac{1}{2}\,S_{20}
                        \,\,=\,\, -i\,\, \frac{1}{2} \sin \left(\frac{\pi N}{4}\right) \nonumber \\
S^J_{(2,0)\widehat{(0,1)}} &=& -\epsilon \,\frac{1}{2} \,e^{-\frac{\epsilon\pi iN}{4}} -\frac{1}{2}\,S_{20}
                        \,\,=\,\, i\,\, \frac{1}{2} \sin \left(\frac{\pi N}{4}\right) \nonumber \\
S^J_{(2,0)\widehat{(1,0)}} &=& \epsilon \,\frac{1}{2} \,e^{\frac{\epsilon\pi iN}{4}} -\frac{1}{2}\,S_{21}
                        \,\,=\,\, i\,\, \frac{1}{2} \sin \left(\frac{\pi N}{4}\right) \nonumber \\
S^J_{(2,0)\widehat{(1,1)}} &=& \epsilon \,\frac{1}{2} \,e^{-\frac{\epsilon\pi iN}{4}} -\frac{1}{2}\,S_{21}
                        \,\,=\,\, -i\,\, \frac{1}{2} \sin \left(\frac{\pi N}{4}\right) \nonumber
\end{eqnarray}
\begin{eqnarray}
S^J_{(2,1)(2,1)} &=& -\frac{1}{2}\,i^N \nonumber \\
S^J_{(2,1)(\phi_0,\phi_1)} &=& -\frac{1}{2} -S_{20}\,S_{21}
                        \,\,=\,\, 0 \nonumber \\
S^J_{(2,1)\widehat{(0,0)}} &=& \epsilon \,\frac{1}{2} \,e^{-\frac{\epsilon\pi iN}{4}} +\frac{1}{2}\,S_{20}
                        \,\,=\,\, -i\,\, \frac{1}{2} \sin \left(\frac{\pi N}{4}\right) \nonumber \\
S^J_{(2,1)\widehat{(0,1)}} &=& \epsilon \,\frac{1}{2} \,e^{\frac{\epsilon\pi iN}{4}} +\frac{1}{2}\,S_{20}
                        \,\,=\,\, i\,\, \frac{1}{2} \sin \left(\frac{\pi N}{4}\right) \nonumber \\
S^J_{(2,1)\widehat{(1,0)}} &=& -\epsilon \,\frac{1}{2} \,e^{-\frac{\epsilon\pi iN}{4}} +\frac{1}{2}\,S_{21}
                        \,\,=\,\, i\,\, \frac{1}{2} \sin \left(\frac{\pi N}{4}\right) \nonumber \\
S^J_{(2,1)\widehat{(1,1)}} &=& -\epsilon \,\frac{1}{2} \,e^{\frac{\epsilon\pi iN}{4}} +\frac{1}{2}\,S_{21}
                        \,\,=\,\, -i\,\, \frac{1}{2} \sin \left(\frac{\pi N}{4}\right) \nonumber
\end{eqnarray}
\begin{eqnarray}
S^J_{(\phi_0,\phi_1)(\phi_0,\phi_1)} &=& \frac{1}{2} -(S_{00}\,S_{11}+S_{01}\,S_{01}) \,\,=\,\, 0 \nonumber \\
S^J_{(\phi_0,\phi_1)\widehat{(0,0)}} &=& -\frac{i}{2} \nonumber \\
S^J_{(\phi_0,\phi_1)\widehat{(0,1)}} &=& \frac{i}{2} \nonumber \\
S^J_{(\phi_0,\phi_1)\widehat{(1,0)}} &=& -\frac{i}{2} \nonumber \\
S^J_{(\phi_0,\phi_1)\widehat{(1,1)}} &=& \frac{i}{2} \nonumber
\end{eqnarray}
\begin{eqnarray}
S^J_{\widehat{(0,0)}\widehat{(0,0)}} &=& \frac{1}{2} \,e^{-\frac{\pi iN}{8}} -\frac{1}{2}\,P_{00} 
                                  \,\,=\,\, -i\,\, \frac{1}{2} \sin \left(\frac{\pi N}{8}\right) \nonumber \\
S^J_{\widehat{(0,0)}\widehat{(0,1)}} &=& -\frac{1}{2} \,e^{\frac{\pi iN}{8}} +\frac{1}{2}\,P_{00} 
                                  \,\,=\,\, -i\,\, \frac{1}{2} \sin \left(\frac{\pi N}{8}\right) \nonumber \\
S^J_{\widehat{(0,0)}\widehat{(1,0)}} &=& \frac{1}{2}\,i \,e^{-\frac{\pi iN}{8}} -\frac{1}{2}\,P_{01} 
                                  \,\,=\,\, i\,\, \frac{1}{2} \cos \left(\frac{\pi N}{8}\right) \nonumber \\
S^J_{\widehat{(0,0)}\widehat{(1,1)}} &=& \frac{1}{2}\,i \,e^{\frac{\pi iN}{8}} +\frac{1}{2}\,P_{01} 
                                  \,\,=\,\, i\,\, \frac{1}{2} \cos \left(\frac{\pi N}{8}\right) \nonumber
\end{eqnarray}
\begin{eqnarray}
S^J_{\widehat{(0,1)}\widehat{(0,1)}} &=& \frac{1}{2} \,e^{-\frac{\pi iN}{8}} -\frac{1}{2}\,P_{00} 
                                  \,\,=\,\, -i\,\, \frac{1}{2} \sin \left(\frac{\pi N}{8}\right) \nonumber \\
S^J_{\widehat{(0,1)}\widehat{(1,0)}} &=& \frac{1}{2}\,i \,e^{\frac{\pi iN}{8}} +\frac{1}{2}\,P_{01} 
                                  \,\,=\,\, i\,\, \frac{1}{2} \cos \left(\frac{\pi N}{8}\right) \nonumber \\
S^J_{\widehat{(0,1)}\widehat{(1,1)}} &=& \frac{1}{2}\,i \,e^{-\frac{\pi iN}{8}} -\frac{1}{2}\,P_{01} 
                                  \,\,=\,\, i\,\, \frac{1}{2} \cos \left(\frac{\pi N}{8}\right) \nonumber
\end{eqnarray}
\begin{eqnarray}
S^J_{\widehat{(1,0)}\widehat{(1,0)}} &=& -\frac{1}{2} \,e^{-\frac{\pi iN}{8}} -\frac{1}{2}\,P_{11} 
                                  \,\,=\,\, i\,\, \frac{1}{2} \sin \left(\frac{\pi N}{8}\right) \nonumber \\
S^J_{\widehat{(1,0)}\widehat{(1,1)}} &=& \frac{1}{2} \,e^{\frac{\pi iN}{8}} +\frac{1}{2}\,P_{11} 
                                  \,\,=\,\, i\,\, \frac{1}{2} \sin \left(\frac{\pi N}{8}\right) \nonumber 
\end{eqnarray}
\begin{eqnarray}
\label{B SJ J=1,0}
S^J_{\widehat{(1,1)}\widehat{(1,1)}} &=& -\frac{1}{2} \,e^{-\frac{\pi iN}{8}} -\frac{1}{2}\,P_{11} 
                                  \,\,=\,\, i\,\, \frac{1}{2} \sin \left(\frac{\pi N}{8}\right) \nonumber \\
&&
\end{eqnarray}

\subsection{$D(n)_1$ series}
The $D(n)_1=SO(N)_1$, $N=2n$, series has central charge $c=\frac{N}{2}$ and four primary fields $\phi_i$ with weight $h_i=0,\frac{N}{16},\frac{1}{2},\frac{N}{16}$ ($i=0,1,2,3$ respectively).
The $S$ matrix is given in table \ref{table S_Dn_1}.
\begin{table}[ht]
\caption{$S$ matrix for $D(n)_1$}
\centering
\begin{tabular}{c|c c c c}
\hline \hline\\
$S_{D(n)_1}$ & $h=0$ &  $h=\frac{N}{16}$ & $h=\frac{1}{2}$ & $h=\frac{N}{16}$ \\ 
\hline &&&\\
$h=0$             & $\frac{1}{2}$        & $\frac{1}{2}$         & $\frac{1}{2}$ &     $\frac{1}{2}$ \\
$h=\frac{N}{16}$  & $\frac{1}{2}$        & $\frac{(-i)^n}{2}$    & $-\frac{1}{2}$ &    $-\frac{(-i)^n}{2}$ \\
$h=\frac{1}{2}$   & $\frac{1}{2}$        & $-\frac{1}{2}$        & $\frac{1}{2}$ &     $-\frac{1}{2}$ \\
$h=\frac{N}{16}$  & $\frac{1}{2}$        & $-\frac{(-i)^n}{2}$   & $-\frac{1}{2}$ &    $\frac{(-i)^n}{2}$
\end{tabular}
\label{table S_Dn_1}
\end{table}

All the four fields of the $D(n)_1$ series are simple currents. 
In the permutation orbifold, they give rise to four integer spin simple currents, namely $(0,0)$, $(0,1)$, $(2,0)$ and $(2,1)$, and to four non-necessarily-integer spin simple current\footnote{For $n$ multiple of $4$, these currents have also integer spin.}, namely $(1,0)$, $(1,1)$, $(3,0)$ and $(3,1)$. We focus here on the first set. Again, the current $(0,0)$ gives a trivial extension. The current $(0,1)$ gives back the tensor product $D(n)_1\times D(n)_1$, with the six off-diagonal fields ($h_{02}=\frac{1}{2}$, $h_{13}=\frac{N}{8}$, $h_{01}=\frac{N}{16}$, $h_{12}=\frac{N}{16}+\frac{1}{2}$, $h_{03}=\frac{N}{16}$, $h_{23}=\frac{N}{16}+\frac{1}{2}$) as fixed points; the $S^J$ matrix is again given by (\ref{SJ_(offdiag.-offdiag.)}).\\
The current $(2,1)$ gives a theory isomorphic to the tensor product. Its fixed points are the fields $(\phi_0,\phi_2)$ ($h=\frac{1}{2}$), $(\phi_1,\phi_3)$ ($h=\frac{N}{8}$), two twisted fields coming from $\phi_1$ (with $h=\frac{N}{16}$ and $\frac{N}{16}+\frac{1}{2}$) and other two from $\phi_3$ (also with $h=\frac{N}{16}$ and $\frac{N}{16}+\frac{1}{2}$), all with same weights as for the off-diagonal fields. The $S^J$ matrix is again equal to the one for $(0,1)$, if the fixed points are ordered suitably according to their weights. \\
As before, more difficult is to derive the $S^J$ matrix for $J=(2,0)$, for which we need (\ref{main formula for f.p. resolution 2}).

\subsubsection{$(D(n)_1)_{\rm perm}$ $S^J$ matrix for $J=(2,0)$}
There are six fixed points for the current $J=(2,0)$ of the permutation orbifold $(D(n)_1)_{\rm perm}$, coming from off-diagonal and twisted fields. They are: $(\phi_0,\phi_2)$ (with $h=\frac{1}{2}$), $(\phi_1,\phi_3)$ (with $h=\frac{N}{8}$), $\widehat{(0,0)}$ ($h=\frac{N}{32}$), $\widehat{(0,1)}$ ($h=\frac{N}{32}+\frac{1}{2}$), $\widehat{(2,0)}$ ($h=\frac{N+8}{32}$) and $\widehat{(2,1)}$ ($h=\frac{N+8}{32}+\frac{1}{2}$).

The $S^J$ matrix can be derived following the same procedure as before. We know $\tilde{S}$ and $S^{BHS}$ and we still have (\ref{main formula for f.p. resolution simplified}). We use the same principle as before to choose the orbit representatives according to their minimal weight. The table in this case is:
\begin{eqnarray}
\bullet & {\rm if \,\,n\,\,is\,\,odd} &  \nonumber \\
(\phi_1,\phi_3) \,\, \longrightarrow&  (0,2N) & \&\qquad (0,14N) \nonumber \\
&&\nonumber \\
\bullet & {\rm if \,\,n\,\,is\,\,even} &  \nonumber \\
(\phi_1,\phi_3) \,\, \longrightarrow&  (1,14N+4) & \&\qquad (3,14N+12) \nonumber \\
&&\\
\bullet & {\rm for \,\, all \,\, n} & \nonumber \\
(\phi_0,\phi_2) \,\, \longrightarrow&  (1,4) & \&\qquad (N-1,16N-4) \nonumber \\
\widehat{(0,0)} \,\, \longrightarrow&  (0,N) & \&\qquad (0,15N) \nonumber \\
\widehat{(0,1)} \,\, \longrightarrow&  (2,15N+8) & \&\qquad (N-2,N-8) \nonumber \\
\widehat{(2,0)} \,\, \longrightarrow&  (N-1,N-4) & \&\qquad (1,15N+4) \nonumber \\
\widehat{(2,1)} \,\, \longrightarrow&  (N-1,15-N) & \&\qquad (1,N+4) \nonumber \\
&&\nonumber
\end{eqnarray}
This fixes our order of ``splitted field 1'' and ``splitted field 2''. We must use fields only from the first set or only from the second set as before. Orbits corresponding to these two splitted fields are conjugate of each other.

The $S^J$ matrix is given below. It depends on the original $S$ and $P$ matrices\footnote{The $P$ matrix for the $D(n)_1$ series is:
\begin{equation}
 P=\left(
\begin{array}{cccc}
\cos\left(\frac{\pi N}{8}\right) & 0 & \sin\left(\frac{\pi N}{8}\right)  & 0 \\
0 & e^{-\frac{i\pi N}{8}}\cos\left(\frac{\pi N}{8}\right)
& 0 & i\,\, e^{-\frac{i\pi N}{8}}\sin\left(\frac{\pi N}{8}\right) \\
\sin\left(\frac{\pi N}{8}\right) & 0 & -\cos\left(\frac{\pi N}{8}\right) & 0 \\
0 &  i\,\, e^{-\frac{i\pi N}{8}}\sin\left(\frac{\pi N}{8}\right)
& 0 & e^{-\frac{i\pi N}{8}}\cos\left(\frac{\pi N}{8}\right)
\end{array}
\right)\,. \nonumber
\end{equation}}
of the $D(n)_1$ theory. We have defined the quantity $r=n {\rm \,\,mod\,\,}2=n-2\left[\frac{n}{2}\right]$, which is $0$ if $n$ is even and $1$ if $n$ is odd. We recall that here $N=2n$. We have checked that it is unitary ($S^J (S^J)^{\dagger}=1$), modular invariant ($(S^J T^J)^3=-1=(SJ)^2$) and gives correct fusion coefficients.

\begin{eqnarray}
S^J_{(\phi_0,\phi_2)(\phi_0,\phi_2)} &=& \frac{1}{2} -(S_{00}\,S_{22}+S_{02}\,S_{02}) \,\,=\,\, 0 \nonumber \\
S^J_{(\phi_0,\phi_2)(\phi_1,\phi_3)} &=& \frac{1}{2} -(S_{01}\,S_{23}+S_{03}\,S_{21}) \,\,=\,\, 0 \nonumber \\
S^J_{(\phi_0,\phi_2)\widehat{(0,0)}} &=& -\frac{i}{2} \nonumber \\
S^J_{(\phi_0,\phi_2)\widehat{(0,1)}} &=& \frac{i}{2} \nonumber \\
S^J_{(\phi_0,\phi_2)\widehat{(2,0)}} &=& -\frac{i}{2} \nonumber \\
S^J_{(\phi_0,\phi_2)\widehat{(2,1)}} &=& \frac{i}{2} \nonumber
\end{eqnarray}
\begin{eqnarray}
S^J_{(\phi_1,\phi_3)(\phi_1,\phi_3)} &=& \frac{1}{2}\,\,i^N -(S_{11}\,S_{33}+S_{13}\,S_{13}) \,\,=\,\, 0 \nonumber \\
S^J_{(\phi_1,\phi_3)\widehat{(0,0)}} &=& -\frac{1}{2}\,\,i^{n+\delta_{r,0}} \nonumber \\
S^J_{(\phi_1,\phi_3)\widehat{(0,1)}} &=& \frac{1}{2}\,\,i^{n+\delta_{r,0}} \nonumber \\
S^J_{(\phi_1,\phi_3)\widehat{(2,0)}} &=& \frac{1}{2}\,\,i^{n+\delta_{r,0}} \nonumber \\
S^J_{(\phi_1,\phi_3)\widehat{(2,1)}} &=& -\frac{1}{2}\,\,i^{n+\delta_{r,0}} \nonumber
\end{eqnarray}
\begin{eqnarray}
S^J_{\widehat{(0,0)}\widehat{(0,0)}} &=& \frac{1}{2}\,\, e^{-\frac{\pi iN}{8}}- \frac{1}{2}\,\,P_{00}                                                    \,\,=\,\, -i\,\,\frac{1}{2}\,\sin\left(\frac{\pi N}{8}\right) \nonumber \\
S^J_{\widehat{(0,0)}\widehat{(0,1)}} &=& -\frac{1}{2}\,\, e^{\frac{\pi iN}{8}}+ \frac{1}{2}\,\,P_{00}                                                    \,\,=\,\, -i\,\,\frac{1}{2}\,\sin\left(\frac{\pi N}{8}\right) \nonumber \\
S^J_{\widehat{(0,0)}\widehat{(2,0)}} &=& \frac{1}{2}\,\,i\,\, e^{-\frac{\pi iN}{8}}- \frac{1}{2}\,\,P_{20}                                                    \,\,=\,\, i\,\,\frac{1}{2}\,\cos\left(\frac{\pi N}{8}\right) \nonumber \\
S^J_{\widehat{(0,0)}\widehat{(2,1)}} &=& \frac{1}{2}\,\,i\,\, e^{\frac{\pi iN}{8}}+ \frac{1}{2}\,\,P_{20}                                                    \,\,=\,\, i\,\,\frac{1}{2}\,\cos\left(\frac{\pi N}{8}\right) \nonumber
\end{eqnarray}
\begin{eqnarray}
S^J_{\widehat{(0,1)}\widehat{(0,1)}} &=& \frac{1}{2}\,\, e^{-\frac{\pi iN}{8}}- \frac{1}{2}\,\,P_{00}                                                    \,\,=\,\, -i\,\,\frac{1}{2}\,\sin\left(\frac{\pi N}{8}\right) \nonumber \\
S^J_{\widehat{(0,1)}\widehat{(2,0)}} &=& \frac{1}{2}\,\,i\,\, e^{\frac{\pi iN}{8}}+ \frac{1}{2}\,\,P_{20}                                                    \,\,=\,\, i\,\,\frac{1}{2}\,\cos\left(\frac{\pi N}{8}\right) \nonumber \\
S^J_{\widehat{(0,1)}\widehat{(2,1)}} &=& \frac{1}{2}\,\,i\,\, e^{-\frac{\pi iN}{8}}- \frac{1}{2}\,\,P_{20}                                                    \,\,=\,\, i\,\,\frac{1}{2}\,\cos\left(\frac{\pi N}{8}\right) \nonumber
\end{eqnarray}
\begin{eqnarray}
S^J_{\widehat{(2,0)}\widehat{(2,0)}} &=& -\frac{1}{2}\,\, e^{-\frac{\pi iN}{8}}- \frac{1}{2}\,\,P_{22}                                                    \,\,=\,\, i\,\,\frac{1}{2}\,\sin\left(\frac{\pi N}{8}\right) \nonumber \\
S^J_{\widehat{(2,0)}\widehat{(2,1)}} &=& \frac{1}{2}\,\, e^{\frac{\pi iN}{8}}+ \frac{1}{2}\,\,P_{22}                                                    \,\,=\,\, i\,\,\frac{1}{2}\,\sin\left(\frac{\pi N}{8}\right) \nonumber
\end{eqnarray}
\begin{eqnarray}
\label{D SJ J=2,0}
S^J_{\widehat{(2,1)}\widehat{(2,1)}} &=& -\frac{1}{2}\,\, e^{-\frac{\pi iN}{8}}- \frac{1}{2}\,\,P_{22}                                                    \,\,=\,\, i\,\,\frac{1}{2}\,\sin\left(\frac{\pi N}{8}\right) \nonumber \\
&&
\end{eqnarray}

\subsection{Comments}
A few comments are in order. \\
First of all, observe that the $S^J$ matrices (\ref{B SJ J=1,0}) and (\ref{D SJ J=2,0}) are purely imaginary.
Secondly, $S^J_{\widehat{(p,\psi)}\widehat{(q,\chi)}}$ does not depend on $\psi$ and $\chi$. \\
Third:
\begin{equation}
S^J_{(\phi_i,\phi_j)\widehat{(p,\psi)}}\,\propto \, e^{i\pi\psi}\,,
\end{equation}
i.e. this entry changes sign as we change $\psi$. \\

\section{Conclusion}
In this paper we have studied the simple current and fixed point structure of permutation orbifolds and we have asked the question of resolving fixed points in these extensions. We did not do it in general but only for the $SU(2)_2$ orbifolds and for the $B(n)_1$ and $D(n)_1$ series. The main results were presented in sections \ref{Fixed point resolution in SU(2)k orbifolds} and \ref{Fixed point resolution in SO(N)1 orbifolds}.

Future directions of research are the following. First we would like to solve the problem in full generality by giving a sensible ansatz for the $S^J$ matrix for an arbitrary CFT. We expect that this ansatz should depend neither on the particular CFT nor on the particular current used in the extension. The results for the special cases considered here give some hints about such a  general
formula, and we hope this will lead us to an educated guess, which can then be checked.

Secondly, the two $SO(N)_1$ series are interesting since they appear in the numerator of the coset CFT defining N=2 minimal models. 
Note that a permutation orbifold of two identical N=2 minimal models is not an N=2 model itself. To impose the word-sheet
supersymmetry, the chiral algebra of the separate factors must be extended by the product of the two supercurrents,
a spin-3 current. Even though our explicit formulas apply only spin-1 currents of permutation orbifolds (apart from the
special case $(0,1)$), this particular spin-3 current is included, since its factor (the supercurrent) originates from the vector representation 
of $SO(2)_1$. The coset construction lifts its conformal weight to $\frac32$, but the fixed point resolution procedure
still applies. Using this extension, we should be able to derive a ``super-BHS" formula for permutation orbifolds
of supersymmetric RCFT's. We intend to study this later on. 

\section*{Acknowledgments}
This research is supported by the Dutch Foundation for Fundamental Research of Matter (FOM)
as part of the program STQG (String Theory and Quantum Gravity, FP 57) and has been partially supported by funding of the Spanish Ministerio de Ciencia e Innovaci\'on, Research Project FPA2008-02968.

\newpage

\begin{appendix}

\section{$S^{J\equiv (2,0)}$ in $SU(2)_2$ orbifolds}
\label{appendix SU2 orbifolds}
In this appendix we would like to give a few details about how to compute the matrix given in table \ref{table S^J2_k=2}. The main tool is the use of formula (\ref{main formula for f.p. resolution 2}), which in this case simplifies into:
\begin{equation}
\label{main formula for f.p. resolution simplified}
 \tilde{S}_{(a,i)(b,j)}=\frac{1}{2}\Big[S^{BHS}_{ab}+ \Psi_i(J)S^{J}_{ab}\Psi_j(J)^{\star}\Big]\,.
\end{equation}
Here, $a$ and $b$ run over the fixed points of $J\equiv (2,0)$, while $i,j=1,2$ refer to the two splitted fields in the extension. Our assignment for the group characters is: $\Psi_1(J)=1,\Psi_2(J)=-1$. It is completely arbitrary and we could have very well made the opposite choice.

$S^{BHS}$ is the orbifold $S$ matrix given in table \ref{table S^BHS_k=2}. The extended $S$ matrix can be derived from the embedding (\ref{SO embedded in SU}) and is given in table \ref{table tildeS^J2_k=2} (the dashed part can be filled out by symmetry). The extending current is $(1,16)$. In details,
\begin{equation}
(SU(2)_2\times SU(2)_2/\mathbb{Z}_2)_{(2,0)}=(SU(3)_1\times U(1)_{48})_{(1,16)}\,.
\end{equation}
Note that $\tilde{S}_{(a,1)(b,j)}=\tilde{S}_{(a,2)(b,j)}^{\star}$, for every $a,\,b,\,j$. Also, observe that we could have interchanged the indices $1\leftrightarrow2$ in the order of the two splitted fields. In deriving the matrix $\tilde{S}$ below, we have chosen the following order for the two splitted fields $(s_1,u_1)$ and $(s_2,u_2)$:
\begin{eqnarray}
(1,0) \,\, \longrightarrow&  (1,46) & \&\qquad (2,2) \nonumber \\
(1,1) \,\, \longrightarrow&  (0,6) & \&\qquad (0,42) \nonumber \\
(\phi_0,\phi_2) \,\, \longrightarrow&  (2,44) & \&\qquad (1,4) \nonumber \\
\widehat{(0,0)} \,\, \longrightarrow&  (0,3) & \&\qquad (0,45) \nonumber \\
\widehat{(0,1)} \,\, \longrightarrow&  (2,5) & \&\qquad (1,43) \nonumber \\
\widehat{(2,0)} \,\, \longrightarrow&  (2,47) & \&\qquad (1,1) \nonumber \\
\widehat{(2,1)} \,\, \longrightarrow&  (0,9) & \&\qquad (0,39) \nonumber
\end{eqnarray}

\begin{table}[ht]
\caption{Fixed point Resolution: Matrix $S^{BHS}$}
\centering
\begin{tabular}{c|c c c c c c c}
\hline \hline\\
$S^{BHS}$ & $(1,0)$ & $(1,1)$ & $(\phi_0,\phi_2)$ & $\widehat{(0,0)}$ & $\widehat{(0,1)}$ & $\widehat{(2,0)}$ & $\widehat{(2,1)}$ \\ 
\hline &&&\\
$(1,0)$             & 0        & 0        & $-2 a$   & $2 b$    & $2 b$    & $-2 b$   & $-2 b$ \\
$(1,1)$             & 0        & 0        & $-2 a$   & $-2 b$   & $-2 b$   & $2 b$    & $2 b$ \\
$(\phi_0,\phi_2)$   & $-2 a$   & $-2 a$   & $2 a$    & 0        & 0        & 0        & 0\\
$\widehat{(0,0)}$   & $2 b$    & $-2 b$   & 0        & $2 c$    & $-2 c$   & $2 d$    & $-2 d$ \\
$\widehat{(0,1)}$   & $2 b$    & $-2 b$   & 0        & $-2 c$   & $2 c$    & $-2 d$   & $2 d$ \\
$\widehat{(2,0)}$   & $-2 b$   & $2 b$    & 0        & $2 d$    & $-2 d$   & $-2 c$   & $2 c$\\
$\widehat{(2,1)}$   & $-2 b$   & $2 b$    & 0        & $-2 d$   & $2 d$    & $2 c$    & $-2 c$
\end{tabular}
\label{table S^BHS_k=2}
\end{table}

\begin{sidewaystable}
\caption{Fixed point Resolution: Matrix $S^{BHS}$}
\centering
\begin{tabular}{c|c c c c c c c c c c c c c c c}
\hline \hline\\
$\tilde{S}_{(a,i)(b,j)}$ & $(1,0)_1$ & $(1,0)_2$ & $(1,1)_1$ & $(1,1)_2$ & $(\phi_0,\phi_2)_1$ & $(\phi_0,\phi_2)_2$ &
 $\widehat{(0,0)}_1$ & $\widehat{(0,0)}_2$ & $\widehat{(0,1)}_1$ & $\widehat{(0,1)}_2$ &
 $\widehat{(2,0)}_1$ & $\widehat{(2,0)}_2$ & $\widehat{(2,1)}_1$ & $\widehat{(2,1)}_2$ \\ 
\hline &&&\\
$(1,0)_1$             & $i a$    & $-i a$   & $i a$    & $-i a$   & $-a$     & $-a$
                      & $b+ib$   & $b-ib$   & $b-ib$   & $b+ib$   & $-b-ib$  & $-b+ib$  & $-b+ib$ & $-b-ib$ &\\
$(1,0)_2$             & $-i a$    & $i a$   & $-i a$    & $i a$   & $-a$     & $-a$
                      & $b-ib$   & $b+ib$   & $b+ib$   & $b-ib$   & $-b+ib$  & $-b-ib$  & $-b-ib$ & $-b+ib$ &\\
$(1,1)_1$             & $i a$    & $-i a$   & $i a$    & $-i a$   & $-a$     & $-a$
                      & $-b-ib$  & $-b+ib$  & $-b+ib$  & $-b-ib$  & $b+ib$   & $b-ib$   & $b-ib$  & $b+ib$  &\\
$(1,1)_2$             & $-i a$    & $i a$   & $-i a$    & $i a$   & $-a$     & $-a$
                      & $-b+ib$  & $-b-ib$  & $-b-ib$  & $-b+ib$  & $b-ib$   & $b+ib$   & $b+ib$  & $b-ib$  &\\
$(\phi_0,\phi_2)_1$   & $-a$     & $-a$     & $-a$     & $-a$     & $a$      & $a$
                      & $i a$    & $-i a$   & $-i a$   & $i a$    & $i a$    & $-i a$   & $-i a$  & $i a$   &\\
$(\phi_0,\phi_2)_2$   & $-a$     & $-a$     & $-a$     & $-a$     & $a$      & $a$
                      & $-i a$   & $i a$    & $i a$    & $-i a$   & $-i a$   & $i a$    & $i a$   & $-i a$  &\\
$\widehat{(0,0)}_1$   & -        & -        & -        & -        & -        & -
                      & $c-id$   & $c+id$   & $-c-id$  & $-c+id$  & $d+ic$   & $d-ic$   & $-d+ic$ & $-d-ic$ &\\
$\widehat{(0,0)}_2$   & -        & -        & -        & -        & -        & -
                      & $c+id$   & $c-id$   & $-c+id$  & $-c-id$  & $d-ic$   & $d+ic$   & $-d-ic$ & $-d+ic$ &\\
$\widehat{(0,1)}_1$   & -        & -        & -        & -        & -        & -        & -       & -
                      & $c-id$   & $c+id$   & $-d+ic$  & $-d-ic$  & $d+ic$ & $d-ic$   &\\
$\widehat{(0,1)}_2$   & -        & -        & -        & -        & -        & -        & -       & -
                      & $c+id$   & $c-id$   & $-d-ic$  & $-d+ic$  & $d-ic$ & $d+ic$   &\\
$\widehat{(2,0)}_1$   & -        & -        & -        & -        & -        & -        & -       & -
                      & -        & -        & $-c+id$  & $-c-id$  & $c+id$ & $c-id$   &\\
$\widehat{(2,0)}_2$   & -        & -        & -        & -        & -        & -        & -       & -
                      & -        & -        & $-c-id$  & $-c+id$  & $c-id$ & $c+id$   &\\
$\widehat{(2,1)}_1$   & -        & -        & -        & -        & -        & -        & -       & -
                      & -        & -        & -        & -        & $-c+id$ & $-c-id$ &\\
$\widehat{(2,1)}_2$   & -        & -        & -        & -        & -        & -        & -       & -
                      & -        & -        & -        & -        & $-c-id$ & $-c+id$ &
\end{tabular}
\label{table tildeS^J2_k=2}
\end{sidewaystable}

At this point, the matrix $S^J$ as given in table (\ref{table S^J2_k=2}) can be derived by subtraction.

\section{Tables for $SU(2)_k$ Orbifold}
\label{Tables for SU(2)k Orbifold}
In order to help the reader follow the first part of the paper, in this appendix we present tables of simple currents and corresponding fixed points and orbits for the $SU(2)_k\otimes SU(2)_k/\mathbb{Z}_2$ orbifold theory for a few values of the level $k$.

In the paper we have analyzed in detail the $k=2$ situation, for which we found all the three $S^J$ matrices. It seems then convenient to list here their orbits corresponding to $(0,1),\,(2,0),\,(2,1)$.

\begin{tabular}{l l l l}
&&& \\
$J\equiv(0,1)$ & \underline{Fixed points} &\phantom{$J\equiv(0,1)$}& \underline{Length-2 orbits} \\
& $(\phi_0,\phi_1)$,\, $h=\frac{3}{16}$   && $\Big( (0,0),(0,1) \Big)$,\, $h=0$ \\
& $(\phi_0,\phi_2)$,\, $h=\frac{1}{2}$    && $\Big( (1,0),(1,1) \Big)$,\, $h=\frac{3}{8}$ \\
& $(\phi_1,\phi_2)$,\, $h=\frac{11}{16}$  && $\Big( (2,0),(2,1) \Big)$,\, $h=1$ \\
&&& \\
\end{tabular}

\begin{tabular}{l l l l}
$J\equiv(2,0)$ & \underline{Fixed points} &\phantom{$J\equiv(2,0)$}& \underline{Length-2 orbits} \\
& $(1,0)$, \,$h=\frac{3}{8}$              && $\Big( (0,0),(2,0) \Big)$,\, $h=0$ \\
& $(1,1)$, \,$h=\frac{11}{8}$             && $\Big( (0,1),(2,1) \Big)$,\, $h=1$ \\
& $(\phi_0,\phi_2)$, \,$h=\frac{1}{2}$\\
& $\widehat{(0,0)}$,\, $h=\frac{3}{32}$ \\
& $\widehat{(0,1)}$,\, $h=\frac{51}{32}$ \\
& $\widehat{(2,0)}$,\, $h=\frac{11}{32}$ \\
& $\widehat{(2,1)}$,\, $h=\frac{27}{32}$ \\
&&& \\
\end{tabular}

\begin{tabular}{l l l l}
$J\equiv(2,1)$ & \underline{Fixed points} &\phantom{$J\equiv(2,1)$}& \underline{Length-2 orbits} \\
& $(\phi_0,\phi_2)$,\, $h=\frac{1}{2}$    && $\Big( (0,0),(2,1) \Big)$,\, $h=0$\\
& $\widehat{(1,0)}$,\, $h=\frac{3}{16}$   && $\Big( (0,1),(2,0) \Big)$,\, $h=1$\\
& $\widehat{(1,1)}$,\, $h=\frac{11}{16}$  && $\Big( (1,0),(1,1) \Big)$,\, $h=\frac{3}{8}$\\
&&& \\
\end{tabular}

It is also interesting to look at larger orbit structures where the pattern discussed in the paper becomes clear. We take e.g. $k=8$. We give here the orbits of the simple currents $(8,0)$ and $(8,1)$. After doubling the fixed points, these give the two extended theories. At this stage, an analogous table for the current $(0,1)$ should be trivial to make.

\begin{tabular}{l l l l}
$J\equiv(8,0)$ & \underline{Fixed points} &\phantom{$J\equiv(8,0)$}& \underline{Length-2 orbits}\\
& $(4,0)$, \,$h=\frac{6}{5}$                 && $\Big( (0,0),(8,0) \Big)$,\, $h=0$\\
& $(4,1)$, \,$h=\frac{6}{5}$                 && $\Big( (0,1),(8,1) \Big)$,\, $h=1$\\
& $(\phi_0,\phi_8)$, \,$h=2$                 && $\Big( (1,0),(7,0) \Big)$,\, $h=\frac{3}{20}$\\
& $(\phi_1,\phi_7)$, \,$h=\frac{33}{20}$     && $\Big( (1,1),(7,1) \Big)$,\, $h=\frac{3}{20}$\\
& $(\phi_2,\phi_6)$, \,$h=\frac{7}{5}$       && $\Big( (2,0),(6,0) \Big)$,\, $h=\frac{2}{5}$\\
& $(\phi_3,\phi_5)$, \,$h=\frac{5}{4}$       && $\Big( (2,1),(6,1) \Big)$,\, $h=\frac{2}{5}$\\
& $\widehat{(0,0)}$,\, $h=\frac{3}{20}$      && $\Big( (3,0),(5,0) \Big)$,\, $h=\frac{3}{4}$\\
& $\widehat{(0,1)}$,\, $h=\frac{13}{20}$     && $\Big( (3,1),(5,1) \Big)$,\, $h=\frac{3}{4}$\\
& $\widehat{(2,0)}$,\, $h=\frac{1}{4}$       && $\Big( (\phi_0,\phi_2),(\phi_6,\phi_8) \Big)$,\, $h=\frac{1}{5}$\\
& $\widehat{(2,1)}$,\, $h=\frac{3}{4}$       && $\Big( (\phi_0,\phi_4),(\phi_4,\phi_8) \Big)$,\, $h=\frac{3}{5}$\\
& $\widehat{(4,0)}$,\, $h=\frac{9}{20}$      && $\Big( (\phi_0,\phi_6),(\phi_2,\phi_8) \Big)$,\, $h=\frac{6}{5}$\\
& $\widehat{(4,1)}$,\, $h=\frac{19}{20}$     && $\Big( (\phi_2,\phi_4),(\phi_4,\phi_6) \Big)$,\, $h=\frac{4}{5}$\\
& $\widehat{(6,0)}$,\, $h=\frac{3}{4}$       && $\Big( (\phi_1,\phi_3),(\phi_5,\phi_7) \Big)$,\, $h=\frac{9}{20}$\\
& $\widehat{(6,1)}$,\, $h=\frac{5}{4}$       && $\Big( (\phi_1,\phi_5),(\phi_3,\phi_7) \Big)$,\, $h=\frac{19}{20}$\\
& $\widehat{(8,0)}$,\, $h=\frac{23}{20}$     &&\\
& $\widehat{(8,1)}$,\, $h=\frac{33}{20}$     &&\\
&&&\\
\end{tabular}

\begin{tabular}{l l l l}
$J\equiv(8,1)$ & \underline{Fixed points} &\phantom{$J\equiv(8,1)$ }& \underline{Length-2 orbits}\\
& $(\phi_0,\phi_8)$, \,$h=2$              && $\Big( (0,0),(8,1) \Big)$,\, $h=0$\\
& $(\phi_1,\phi_7)$, \,$h=\frac{33}{20}$  && $\Big( (0,1),(8,0) \Big)$,\, $h=1$\\
& $(\phi_2,\phi_6)$, \,$h=\frac{7}{5}$    && $\Big( (1,0),(7,1) \Big)$,\, $h=\frac{3}{20}$\\
& $(\phi_3,\phi_5)$, \,$h=\frac{5}{4}$    && $\Big( (1,1),(7,0) \Big)$,\, $h=\frac{3}{20}$\\
& $\widehat{(1,0)}$,\, $h=\frac{3}{16}$   && $\Big( (2,0),(6,1) \Big)$,\, $h=\frac{2}{5}$\\
& $\widehat{(1,1)}$,\, $h=\frac{11}{16}$  && $\Big( (2,1),(6,0) \Big)$,\, $h=\frac{2}{5}$\\
& $\widehat{(3,0)}$,\, $h=\frac{27}{80}$  && $\Big( (3,0),(5,1) \Big)$,\, $h=\frac{3}{4}$\\
& $\widehat{(3,1)}$,\, $h=\frac{67}{80}$  && $\Big( (3,1),(5,0) \Big)$,\, $h=\frac{3}{4}$\\
& $\widehat{(5,0)}$,\, $h=\frac{47}{80}$  && $\Big( (4,0),(4,1) \Big)$,\, $h=\frac{6}{5}$\\
& $\widehat{(5,1)}$,\, $h=\frac{87}{80}$  && $\Big( (\phi_0,\phi_2),(\phi_6,\phi_8) \Big)$,\, $h=\frac{1}{5}$\\
& $\widehat{(7,0)}$,\, $h=\frac{15}{16}$  && $\Big( (\phi_0,\phi_4),(\phi_4,\phi_8) \Big)$,\, $h=\frac{3}{5}$\\
& $\widehat{(7,1)}$,\, $h=\frac{23}{16}$  && $\Big( (\phi_0,\phi_6),(\phi_2,\phi_8) \Big)$,\, $h=\frac{6}{5}$\\
&                                         && $\Big( (\phi_2,\phi_4),(\phi_4,\phi_6) \Big)$,\, $h=\frac{4}{5}$\\
&                                         && $\Big( (\phi_1,\phi_3),(\phi_5,\phi_7) \Big)$,\, $h=\frac{9}{20}$\\
&                                         && $\Big( (\phi_1,\phi_5),(\phi_3,\phi_7) \Big)$,\, $h=\frac{19}{20}$\\
\end{tabular}

\end{appendix}

\end{document}